\documentclass[a4paper,aps,superscriptaddress,floatfix,nofootinbib,twocolumn,bibtex]{revtex4}

\usepackage{bbold}
\usepackage{bbm}
\usepackage[pdftex]{graphicx}
\usepackage{latexsym,amsmath,verbatim}
\usepackage{color}
\usepackage{rotating}
\usepackage{verbatim}
\usepackage{multirow}
\usepackage[english]{babel}
\usepackage{comment}
\usepackage[final]{pdfpages}

\usepackage[caption=false]{subfig}

\usepackage{morefloats}

\usepackage{etoolbox}

\usepackage{amsmath,amssymb}

\begin{document}

\title{Influence maximization in noisy networks}

\author{\c{S}irag Erkol}
\affiliation{Center for Complex Networks and Systems Research, School
  of Informatics, Computing, and Engineering, Indiana University, Bloomington,
  Indiana 47408, USA}

\author{Ali Faqeeh}
\affiliation{Center for Complex Networks and Systems Research, School
  of Informatics, Computing, and Engineering, Indiana University, Bloomington,
  Indiana 47408, USA}
\affiliation{MACSI, Department of Mathematics \& Statistics, University of Limerick, Limerick, Ireland}

\author{Filippo Radicchi}
\affiliation{Center for Complex Networks and Systems Research, School
  of Informatics, Computing, and Engineering, Indiana University, Bloomington,
  Indiana 47408, USA}
\email{filiradi@indiana.edu}

\begin{abstract}
We consider the problem of identifying the most influential
nodes for a spreading process on a network when prior
knowledge about structure and dynamics of the
system is incomplete or erroneous. Specifically, we perform a numerical
analysis where the set of
top  spreaders  is determined on the basis of
prior information that is artificially altered by a certain level of noise.
We then measure the optimality of the chosen set
by measuring its spreading impact in the true system.
Whereas we find that the identification of top spreaders is
optimal when prior knowledge is complete and free
of mistakes, we also find that the quality of the top spreaders
identified using noisy information doesn't necessarily decrease
as the noise level increases. For instance, we show that it is
generally possible to compensate for erroneous information about dynamical parameters
by adding synthetic errors in the structure of the network. Further, we show that,
in some dynamical regimes, even completely losing prior knowledge on
network structure may be better than relying on certain but incomplete
information.
\end{abstract}

\maketitle

\section{Introduction}

In a social network where an opinion diffuses according
to an irreversible spreading process,
a fundamentally important role for the
ultimate success of the opinion
is played by the nodes that act as initiators or seeds of the
spreading process. For example, the popularity of
memes in social media is often determined
by just a few early adopters~\cite{ratkiewicz2011truthy}.
Given the high sensitivity of the outcome of
spreading processes to initial configurations, a very
interesting problem regards the
identification of the initial configuration, among the many possible,
that maximizes the extent of
diffusion.  The problem is traditionally named as
influence maximization. It has been first considered
by Domingos and Richardson~\cite{Domingos01},
and slightly later generalized by
Kempe {\it et al.}~\cite{Kempe03}.
Roughly speaking, influence maximization consists in an optimization problem
based on a few assumptions and subjected to one constraint.
The function that one wants to maximize is the size of the outbreak,
i.e., the number of nodes that will end up acquiring the opinion that
is diffusing in the system. The assumptions in the formulation
of the problem regard the structure of the network
and the type of spreading that is taking place on the network.
This information is generally assumed as a prior knowledge, and it
is actively used for finding solutions to the optimization problem.
The only constraint used in the optimization problem is
the number of seeds. Only initial configurations consisting
of a given number of active nodes
are considered as potential solutions to the optimization problem.

As most optimization problems, influence maximization
is NP-complete~\cite{Kempe03}. Exact solutions
can be found only in very small networks.
Suboptimal solutions can be achieved with
approximated or greedy
optimization techniques~\cite{Kempe03, Chen09}.
These approaches are generally effective, but they are designed for
the analysis of small to medium networks.
The identification of influential spreaders in
large networks is allowed only through the use
of heuristic techniques where dynamics is {\it de facto}
neglected, and the solution to the optimization problem is
approximated relying on network
centrality metrics~\cite{Kitsak10, Bauer12, klemm2012measure,
Chen12, Chen13,
PhysRevE.90.032812, Morone15}. This approach finds
its rationale in interpreting the high sensitivity of the outcome of
a spreading process to the initial conditions
as a consequence of the heterogeneity of
the underlying network. However, geometry alone
is not enough to provide a sufficiently
accurate description of the state of a dynamical system
running on a network~\cite{radicchi2017maximum}.
The identity of the most influential
nodes in a network generally changes from type to type of
spreading process, and, even for the same type of process, it may
depend on its dynamical regime~\cite{radicchi2017fundamental}.

Most of the studies we mentioned above
rely on one strong assumption: prior
knowledge of system structure and dynamics is
complete and free of errors. When dealing with a real application
of influence maximization, we should however recognize that this
assumption is at least optimistic.~The presence/absence of a connection in a
social network is generally established from the result
of some experimental observation, and it is therefore
potentially  affected by experimental errors~\cite{Newman17}.
Similarly, we may be aware of the type of
process that drives spreading, but we may be unsure
about the exact value of the rates at which spreading occurs.
There are techniques that allow to accurately estimate spreading rates
from empirical observations of spreading
events~\cite{saito2008prediction}.
However, these techniques rely on the
assumption that structural
information is complete and free of mistakes. Further, in
influence maximization, one aims at controlling the
fate of a future or ongoing spreading process, so posterior
estimates of the rates are not very helpful.

Several previous
studies have considered the
reliability of network centrality metrics when computed from
noisy or incomplete structural
information~\cite{dall2005statistical, Leskovec2006}.
In the context of influence maximization, we are aware
of previous tests of robustness of some centrality metrics
in noisy structural data~\cite{Kitsak10}. However,
to the best of our knowledge, there are no previous studies that
attempted to understand how incomplete or
erroneous information about both structure and
dynamics affects our ability to solve the problem
of influence maximization. Please note that we may
naively expect that noise doesn't dramatically modify
the overall trend of a geometric centrality metric, as it was
shown in Ref.~\cite{Kitsak10}. However,
the distortions that noise can create in the solutions of
an optimization problem such as influence maximization
are far less predictable. The current paper aims at filling
this gap of knowledge.

\section{Methods}

\subsection{Network structure}

We assume that the network where the spreading
process takes
place and where we aim at solving
the problem of influence maximization
is given by $N$ total nodes and $M$ edges.
The network is unweighted and undirected.
Structural information about the network is
fully specified by the adjacency matrix $A$,
whose generic element
$A_{ij} = A_{ji} =1$ if nodes $i$ and $j$ are connected,
and  $A_{ij} = A_{ji} =0$, otherwise.
Note that, in the spreading process, only the state
associated to the nodes of the network can change.
Edges do not have states that evolve in time, but serve
only as static media for spreading.

\subsection{Spreading dynamics}

In this paper, we focus our attention on
a spreading model that is very popular
in studies about the identification of influential spreaders
in networks: the Independent Cascade Model
(ICM)~\cite{Kempe03}. The ICM is
very similar to the traditional
Susceptible-Infected-Recovered (SIR)
model~\cite{PastorSatorras15}.
During ICM dynamics, nodes in the network can be found in three
different states: S, I, or R. Generally, in an initial configuration
of the dynamics, all nodes
are set in state S, except for a subset of seeds $Q$ that are
set in state I. At each discrete stage of the
dynamics, two rules are applied in sequence: (i) every
node in state I infects, with probability $p$, each of its neighbors in state
S; (ii) all nodes in state I, that attempted
to infect their neighbors at step (i), recover and change their state
from I to R. Rules (i) and (ii) are iterated until no infected nodes
are longer present in the network. The size of the outbreak
is given by the total number of nodes that are in state R
at the end of the dynamics. This number is a stochastic variable
that may change its value from instance to instance of the
model. Observed values depend on the network
structure encoded by the adjacency matrix $A$,
the value of the spreading probability $p$, and the set of seeds $Q$
that initiated the spreading process. In our numerical study, we measure
the spreading intensity for given $A$, $p$, and $Q$ in terms of the average
value of outbreak size, namely $O$,
over $20$ independent
simulations of the spreading process.

\subsection{Influence maximization}

Solving the problem of influence maximization
for a spreading process on a network
means finding the  set of seeds that maximizes
the average value of the outbreak size $O$. The
maximization is performed for a fixed size $|Q|$ of the
set of seeds.  The optimization relies on
prior knowledge about the structure of the network and the dynamical
rules of the
spreading process.~Information on the structure is provided by the adjacency matrix $A$.
Information about the dynamics consists in knowing that spreading is
regulated by the ICM and the value of the probability of spreading is
$p$. Even with full knowledge of
system structure and dynamics,
identifying the set of optimal seeds
is a NP-complete problem, and thus
exact solutions are achievable only in extremely small networks~\cite{Kempe03}.
Greedy optimization, however,
allows to provide suboptimal solutions that are granted to be
within $63\%$ of the optimum~\cite{Kempe03}.
This is the consequence of the fact that the size of the outbreak $O$
is a submodular function of the set of
seeds~\cite{Kempe03, Nemhauser1978}.
According to greedy optimization,
the set of influential spreaders $Q$ is constructed sequentially
by adding one node at a time. At every stage,
the best node to be added to the set of seeds is the one that
leads to the maximal value of $O$,
in a spreading process initiated by all nodes that
are already part of the seed set plus the node under consideration.
In the original version of the algorithm by Kempe {\it et
  al.}~\cite{Kempe03}, greedy steps rely on direct
simulations of the dynamical process. This algorithm is very general,
and can be used for other types of dynamical models, not just the ICM.
For the specific case of the ICM, the greedy
algorithm can be further speeded up by taking
advantage of the mapping between static properties of the SIR and bond
percolation~\cite{grassberger1983critical}.  The mapping allows to
use configurations of the percolation model to infer
information about final configurations arising
from ICM dynamics. Each node is associated with a score whose value
  is proportional to the average size of independent clusters (i.e.,
  clusters that do not contain previously identified seed nodes) which
  the node belongs to. Now the set $Q$ is constructed by adding the nodes to it, one by one, starting from the nodes with the highest score.
This method was designed, implemented and validated by Chen
{\it et al.}~\cite{Chen09}.  The results of the current paper
are based on our re-implementation of the algorithm by Chen
{\it et al.}  We rely on $R = 1,000$ independent
simulations of the bond percolation model to compute the scores
associated with the nodes. Please note that as long as 
$R$ is a finite number, scores associated with nodes are subjected to finite-size
  fluctuations. As a consequence, the solution to the optimization
  problem provided by the algorithm by Chen {\it et al.} may
depend on the specific set of bond percolation simulations considered.
As the results of the SM show, we verified that, while the identity of the nodes in the
solutions provided by the algorithm is not always the same,
the average sizes $O$ of the outbreak associated with those solutions
are very similar. This finding suggests that the exact
  detection of the set of top
  spreaders is statistically irrelevant for the
outcome of the spreading process, in the sense that
there are many nearly-optimal solutions to the 
problem of influence maximization.

\subsection{Modeling errors in system dynamics and structure}
The process of selection of top spreaders relies
on prior knowledge about the structure
of the underlying network
and the details of the dynamical process.~This means that the set $Q$ depends on the information at our disposal
regarding the structure of the network, i.e., the adjacency matrix
$A$. $Q$ also depends on our prior knowledge about the
dynamical process that is taking place on the
network, that is the ICM model with spreading probability equal to $p$.
It is common practice to assume prior information complete and exact.~This is equivalent to assuming that the inputs $A$ and $p$ of the
algorithm for the identification of top spreaders are equal to their
true values, namely $A_{true}$ and $p_{true}$, respectively.
However, in practical situations, this may not be the case.
Prior knowledge may be affected by errors,
so that the actual information
used to solve the problem of influence maximization is given by
$A_{err}$ and $p_{err}$, respectively. There are potentially many
different ways to model errors that deteriorate
dynamical and structural information of the
system. Here, we opt for simple, yet realistic, models.

Errors that affect our prior knowledge of the spreading dynamics
are simply obtained by setting $p_{err} \neq p_{true}$. Essentially, we assume
to know that spreading occurs following the rules
of the ICM, but we pretend that we don't know the exact value of
the spreading probability. Instead of using raw values
for the spreading probabilities,  we rescale them
as  $\phi_{true} = p_{true} / p_c$ and
$\phi_{err} = p_{err} / p_c$, where $p_c$ is the critical
value of the spreading probability for the ICM
model on the true network $A_{true}$. This transformation is used only
to simplify the presentation of the results. It allows us in fact to
use the same reference value for all networks to
distinguish between different dynamical
regimes of spreading. The critical value of the
spreading probability is computed directly from numerical simulations where
ICM is initiated from a randomly selected seed node~\cite{Radicchi16}.
Depending on whether $\phi_{true}$ is larger, equal or smaller than
one, we say that the network is respectively in the supercritical,
critical or subcritical regime of spreading. Similarly, the value of
$\phi_{err}$ tells us what type of dynamical
regime is hypothesized for the selection of influential spreaders.
From previous studies, we know that the identity of the nodes that
populate the sets of top spreaders is highly dependent
on the regime of spreading~\cite{radicchi2017fundamental}.
We expect therefore that the error in the estimation of
the spreading probability may strongly affect our
ability to properly predict top spreaders.

Errors in the structure of the network are generated artificially
according to a model similar to the one considered in
Ref.~\cite{Newman17}. The total number of nodes is unaffected by
noise, so that we can indicate it as $N$.
Errors happen at the level of pairs of nodes.
This means that the  total number of edges $M_{err}$ in the
altered structure generally differs from $M_{true}$, i.e.,  the number of
edges in the true network. We consider two potential sources of
errors. The first source is responsible for making true edges
disappear from our prior knowledge of the network. Specifically, given
the true  adjacency matrix $A_{true}$, every pair
of connected nodes is disconnected with probability $0 \leq
\epsilon_{del} \leq 1$ in $A_{err}$.  The number of true edges that
are deleted equals zero for
$\epsilon_{del} = 0$, and equals $M_{true}$
for $\epsilon_{del}=1$.  The second source of error
generates false edges. Every pair of nodes that is not connected according
 to $A_{true}$ appears as connected
in $A_{err}$ with probability
$\epsilon_{add} \, M_{true} / [N(N-1)/2-M_{true}]$, where $0 \leq
\epsilon_{add} \leq 1$.
For $\epsilon_{add} = 0$, no false edges are
added to the true network.  For $\epsilon_{add} = 1$ instead,
the expected total number of false edges equals $M_{true}$.
The definition of the noise parameters $\epsilon_{del}$ and
  $\epsilon_{add}$ are such that both parameters are confined in the
  range $[0,1]$, and their maximum values correspond to an expected
  alteration (i.e., deletion or addition) of $100\%$ of the true number of edges.

Please note that removing true edges is generally not
the inverse operation of adding false edges.
For instance, even the addition of a very small number of edges among pairs of
non-connected nodes is able to decrease substantially the average path
length of graphs that do not originally satisfy the
small-world property~\cite{watts1998collective}.
For example, in networks with strong spatial embedding,
as some of those we analyze in this paper, random edges
likely behave as shortcuts
between spatially far regions of the system.
On the other hand, the removal
of a small fraction of true edges
doesn't change dramatically the average
path length of the graph. Also, we do not expect the two
sources of structural errors to be equally likely in real networks.
Their proportion may depend strongly on the type
of network considered, and on the way the network is actually
constructed from empirical observations~\cite{Newman17}.
As the two sources of structural errors cannot be treated on the same
footing,  in our analysis we always consider them separately, i.e.,
we always work with the condition $\epsilon_{del} \,
\epsilon_{add} =0$ satisfied.

We stress that our choice for the noise model affecting prior
  structural information is heavily inspired by
  Ref.~\cite{Newman17}. We find the model simple enough, yet able to
naturally describe sources of uncertainty in empirically constructed
social networks. Alternative models of
structural noise could be considered and studied using the same exact
methods as those described here. For example,
a model consisting in shuffling true edges with a
certain probability would provide a way to introduce
structural noise without
altering the degree of the nodes in
the network. In general, we believe that the choice of the noise model
should depend on the specific question that one wants to address,
or the specific system that one is considering. The set of questions
that we are considering here can be fully addressed by the particular choice
we made.

\subsection{Measuring performance}
Given the inputs $A_{err}$
and $\phi_{err}$, we make use of
the algorithm by Chen {\it et al.}
to identify the set
$Q_{err}$ of top spreaders. As the
algorithm for the identification of top
spreaders is based on a finite number of numerical simulations
  and the output of the algorithm is subjected to finite-size fluctuations,
we apply the algorithm $V = 10$ times to find $V$ potentially
different sets $Q_{err}$. For each of them, we use numerical
simulations of the
ICM model relying on  $A_{true}$ and $p_{true}$ to evaluate
performance.
In particular, as the algorithm by Chen {\it et al.}~naturally
ranks nodes  according to the order in which they
are added to the set of top spreaders, we explicitly use this
information to quantify the performance of the set $Q_{err}$
as follows.  The size of the set $Q_{err}$ is indicated with $|Q_{err}|$. Nodes have
been added to the set according to the sequence $q_1, q_2, \ldots,
q_{|Q_{err}|}$. Define $Q_{err}^{(r)} = \cup_{v=1}^{r} q_v$ as the
set of nodes with rank up to $r$. Please note that, by definition,
$Q_{err}^{(|Q_{err}|)} \equiv Q_{err}$.  The overall performance of the
set $Q_{err}$ is computed according to the equation
\begin{equation}
P (Q_{err}) =  \frac{1}{N \, |Q_{err}|}  \,  \sum_{r=1}^{|Q_{err}|} \,
O(Q^{(r)}_{err}) \; ,
\label{eq:performance}
\end{equation}
where $O(Q^{(r)}_{err})$ is the average size of the
outbreak in the ICM when the set of seeds is
given by $Q^{(r)}_{err}$.  For given $A_{err}$,
the overall performance  is finally obtained by taking the average over
all $V$ realizations of the sets of top spreaders. To account for the
stochastic nature of structural noise, we repeat the entire procedure
on $G = 10$ instances of $A_{err}$ and quantity the spreading performance
of top spreaders as the average value over these independent
realizations.
Please note that the sum on the r.h.s. of
Eq.~(\ref{eq:performance}) allows us to estimate not only the
overall performance of the set $Q_{err}$, but also the way the set is
constructed. The pre-factor appearing in the r.h.s. of
Eq.~(\ref{eq:performance}) is used only to confine values of
our performance metric to the interval $[0,1]$.
Our focus here is not on measuring the
effectiveness of the algorithm used to determine the spreaders, rather
on the importance that prior information has in the selection of the
top spreaders. Other metrics could be used in place of the one
defined in Eq.~(\ref{eq:performance}).  For instance, metrics that
consider the identity of the nodes in $Q_{err}$ with respect to those
found in the true set of top spreaders. We believe, however, that this
second type of metric may be misleading as the difference in terms
of outbreak size between the optimal solution and a slightly-less
optimal solution may be very small despite a high dissimilarity in
terms of the nodes that define the two solutions.
In the SM, we actually verified that the identity
  of the nodes in the set $Q_{err}$ may be sensitive to the choice of
  the noise parameters and the specific run of the identification algorithm;
instead, the size of the outbreak $O$ is not much affected by
fluctuations.

\section{Results}

\begin{figure}[!htb]
\includegraphics[width=0.45\textwidth]{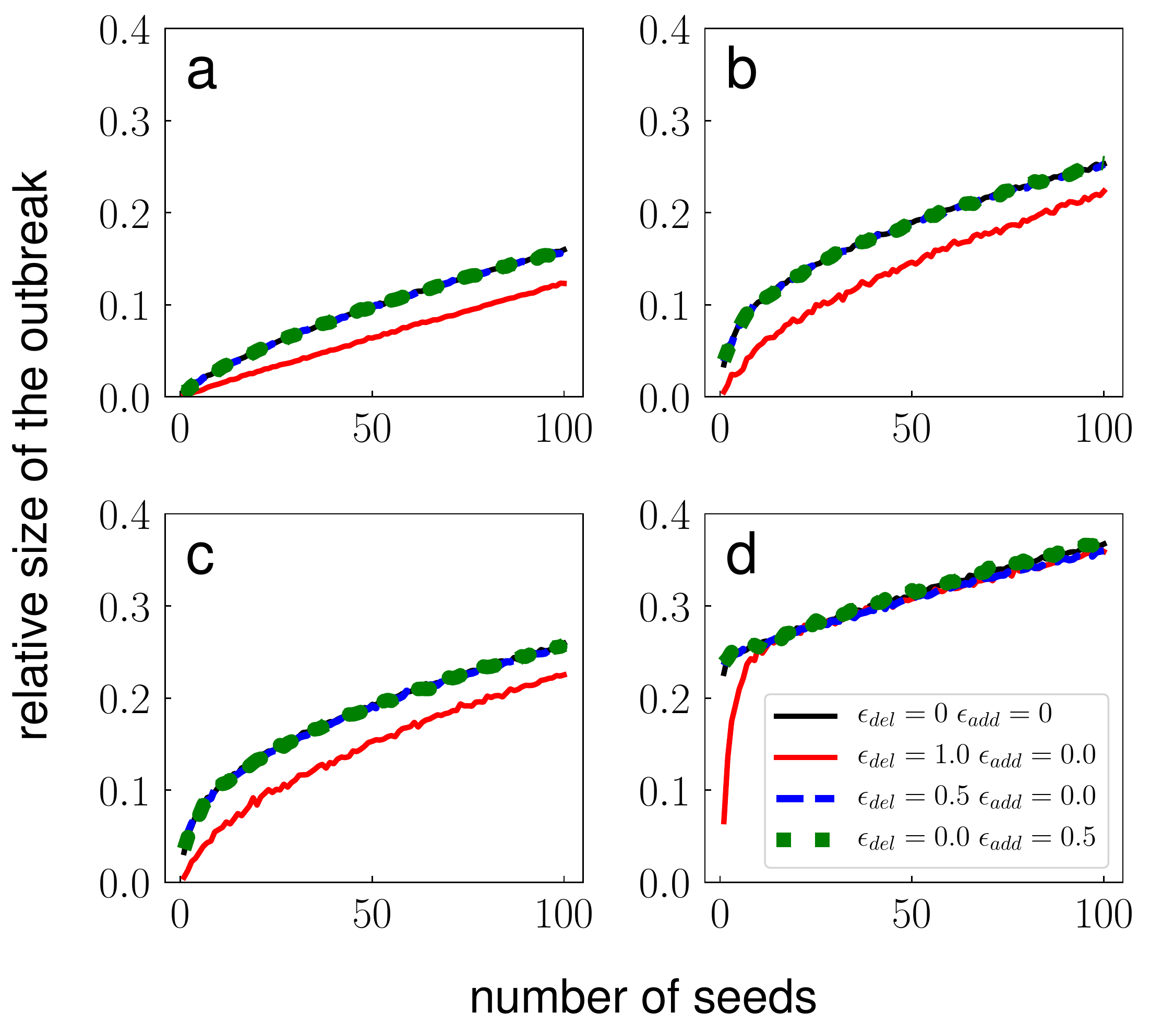}
\caption{Influence maximization in presence of structural and
  dynamical noise. The true network structure analyzed here is
given by the email communication
network of Ref.~\cite{PhysRevE.68.065103}. (a) Relative size
of the outbreak $O/N$ as a
function of the number of seeds found by greedy optimization.~The true spreading probability of the ICM is such that $\phi_{true} =
0.5$. Prior dynamical knowledge used by the greedy algorithm is not
affected by noise, i.e., $\phi_{err} = \phi_{true} = 0.5$. The different curves correspond to different level of
noise that affect prior structural information.~We consider various
combinations of the parameters  $\epsilon_{del}$
and $\epsilon_{add}$. (b) Same as in panel a, but for $\phi_{true} =
1.0$ and $\phi_{err} =0.5$. (c) Same as in panel a, but for $\phi_{true} =
\phi_{err} =1.0$. (b) Same as in panel a, but for $\phi_{true} =
1.5$ and $\phi_{err} =1.0$.
}
\label{fig:1}
\end{figure}

In Figure~\ref{fig:1}, we display results obtained for the
email communication network originally considered in
Ref.~\cite{PhysRevE.68.065103}. In the majority of the cases,
the performance of
the set of top spreaders appears robust against structural noise. However,
as Figure~\ref{fig:1} shows, the overall ability to properly select
top spreaders may be seriously affected by the level of noise
associated with
prior information both at the structural and the dynamical levels.
Major issues seem to arise when $\phi_{true} > 1$, but
$\phi_{err} < 1$ (see Fig.~\ref{fig:1}d).

\begin{figure}[!htb]
\includegraphics[width=0.45\textwidth]{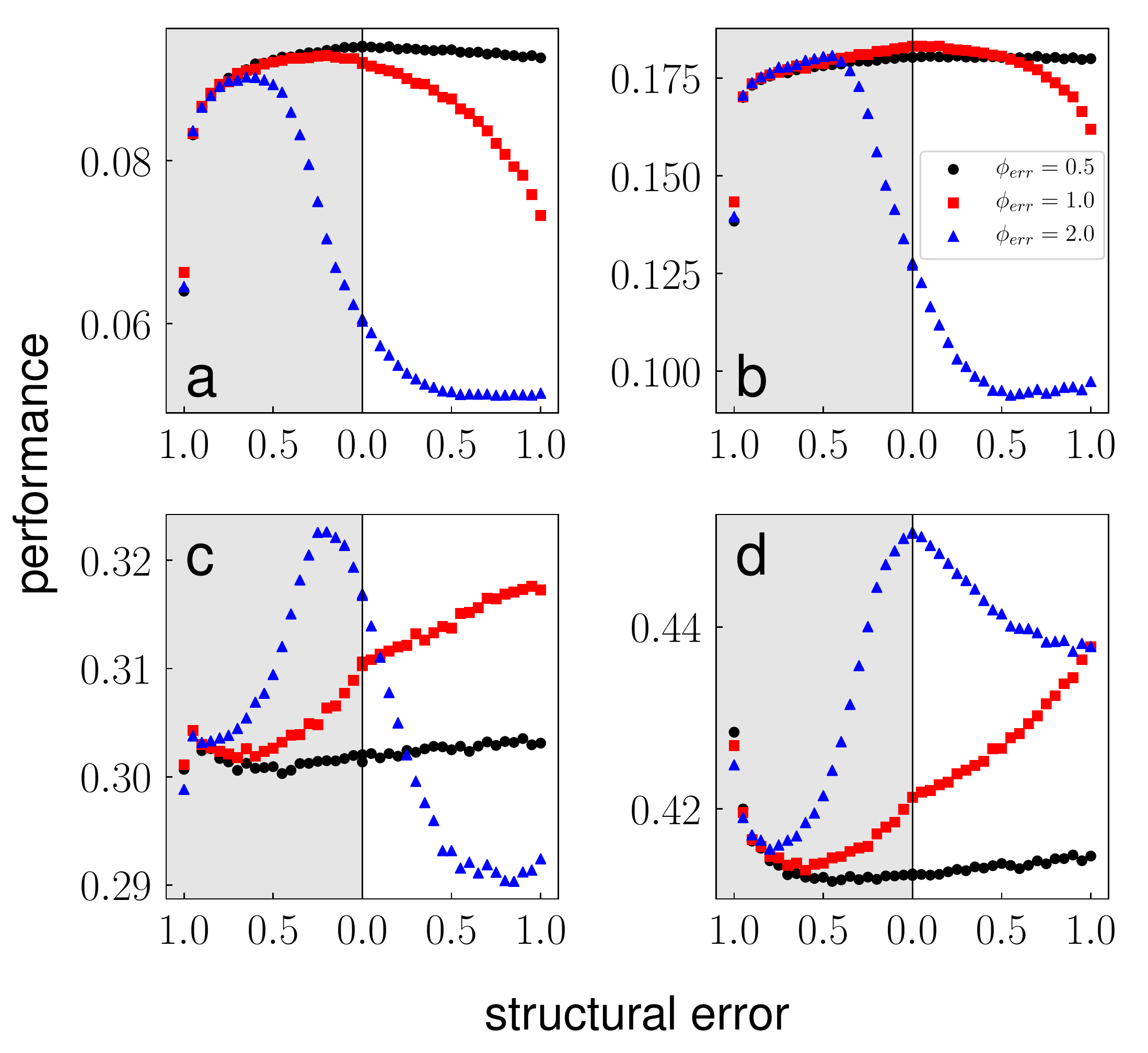}
\caption{Performance of the top spreaders in the presence of structural and
  dynamical noise.~We consider the same network as in Fig.~\ref{fig:1}. (a) We compute
Eq.~(\ref{eq:performance}) for the set of top spreaders of size $|Q_{err}|
= 100$, and we plot the value of the performance as a function of the
noise level in prior structural information.~Performance is
measured for $\phi_{true} = 0.5$. The shaded part of the plot
serves to report results valid for $0 \leq \epsilon_{del} \leq 1 $ and $\epsilon_{add}
=0$. The non-shaded part of the graph represents instead  results for
$0 \leq \epsilon_{add} \leq 1 $ and $\epsilon_{del}
=0$. (b) Same as in panel a, but for
$\phi_{true} = 1.0$. (c) Same as in panel a, but for
$\phi_{true} = 1.5$. (d) Same as in panel a, but for
$\phi_{true} = 2.0$.
}
\label{fig:2}
\end{figure}

To better characterize the observed trend,
we set
$|Q_{err}| = 100$ and quantify the performance defined in
Eq.~(\ref{eq:performance}) for different levels of noise.
The results of this analysis are presented
in Figure~\ref{fig:2}.~The various
panels of the figure
refer to different dynamical regimes identified by different values of
$\phi_{true}$.~In every panel, we present three curves, each
representing a specific value of $\phi_{err}$.~Each curve stands for spreading performance
$P(Q_{err})$ as a function of the structural noise
parameters $\epsilon_{del}$ and $\epsilon_{add}$.~Please consider that, although the two
sources of structural noise are never considered
active simultaneously, we
present them in the same plot for the sake of
compactness.~The general observed behavior can be summarized as follows.
Maximal performance is reached at $\epsilon_{del} = \epsilon_{add} =
0$, only for $\phi_{err} = \phi_{true}$.
The two sources of structural noise affect the choice of the top
spreaders
differently.
Consider first the case $\epsilon_{del} =0$, but $0 \leq \epsilon_{add}
\leq 1$. Roughly, we see that $P(Q_{err})$ is a monotonic function of
$\epsilon_{add}$, decreasing if $\phi_{err} \geq \phi_{true}$, and
increasing, otherwise.
In the region $\epsilon_{add} = 0$ and $0 \leq \epsilon_{del}
\leq 1$, instead, $P(Q_{err})$ is not a monotonic
function of the structural error. Further, the trend changes
depending on whether the system is in the subcritical
or supercritical regime: if $\phi_{true} \leq 1$, $P(Q_{err})$
is concave;  if $\phi_{true} \geq 1$, $P(Q_{err})$
is convex. In this second regime, it becomes possible to
obtain higher performance by adding further structural mistakes.
If $\phi_{err} < \phi_{true}$, best performance is achieved for
$\epsilon_{del} =1$, essentially using no prior knowledge of
the network structure, i.e. seeds are sampled at random.

\begin{figure}[!htb]
\includegraphics[width=0.45\textwidth]{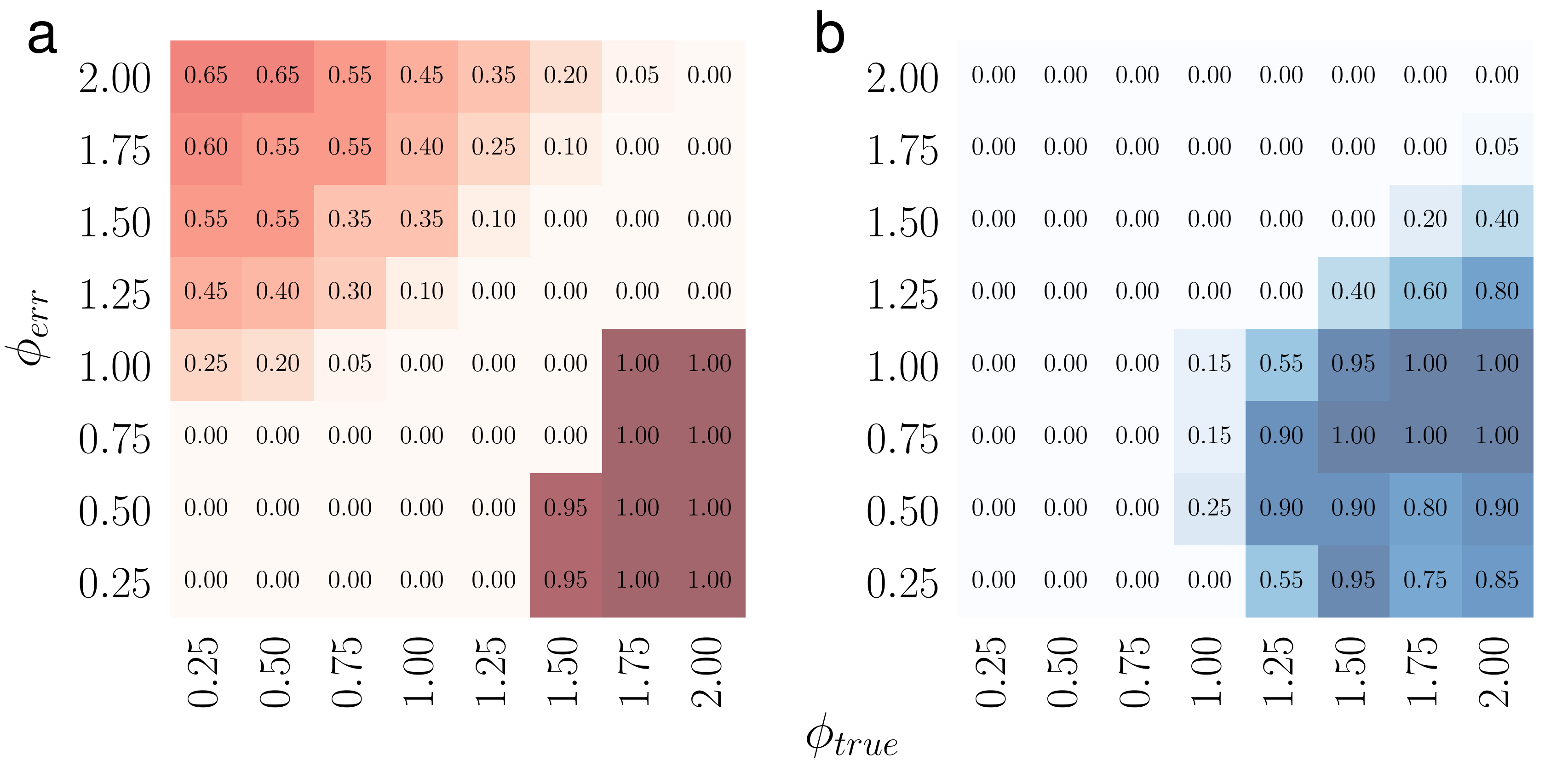}
\caption{Best values of the structural errors in the presence of dynamical
  uncertainty. We consider the same network as in Fig.~\ref{fig:1}. (a) We set $\epsilon_{add}
= 0$, and, for given dynamical parameters $\phi_{true}$ and $\phi_{err}$,
we determine $\hat{\epsilon}_{del} $, i.e., the value of
the error parameter $\epsilon_{del} $ that
leads to the maximum performance in the prediction of top spreaders.
Best estimates of $\hat{\epsilon}_{del} $  are reported in the cells
of the table.
The intensity of the background color is proportional
to the value of $\hat{\epsilon}_{del} $. (b) Same as in panel a, but
for the other source of structural error.~We set here $\epsilon_{del}
= 0$ and focus on $\hat{\epsilon}_{add} $, i.e., the value of
the error parameter $\epsilon_{add} $ that
allows to identify the best performing set of top spreaders.
}
\label{fig:3}
\end{figure}

We analyzed the phenomenon systematically.
For different combinations $\phi_{true}$ and $\phi_{err}$,
we computed $\hat{\epsilon}_{del}$ and $\hat{\epsilon}_{add}$, i.e.,
the values of the parameters of structural noise where $P(Q_{err})$
reaches its maximum.  As Figure~\ref{fig:3}a shows,
for $\phi_{err} \simeq \phi_{true}$, $\hat{\epsilon}_{del} \simeq 0$.
However, as soon as the error on the dynamical  parameter
increases, this error may be compensated by
making further mistakes in the structure.
When structural noise is allowed through the
addition of edges instead, noise is helpful only in the
supercritical regime (Fig.~\ref{fig:3}b).

\begin{figure}[!b]
\includegraphics[width=0.45\textwidth]{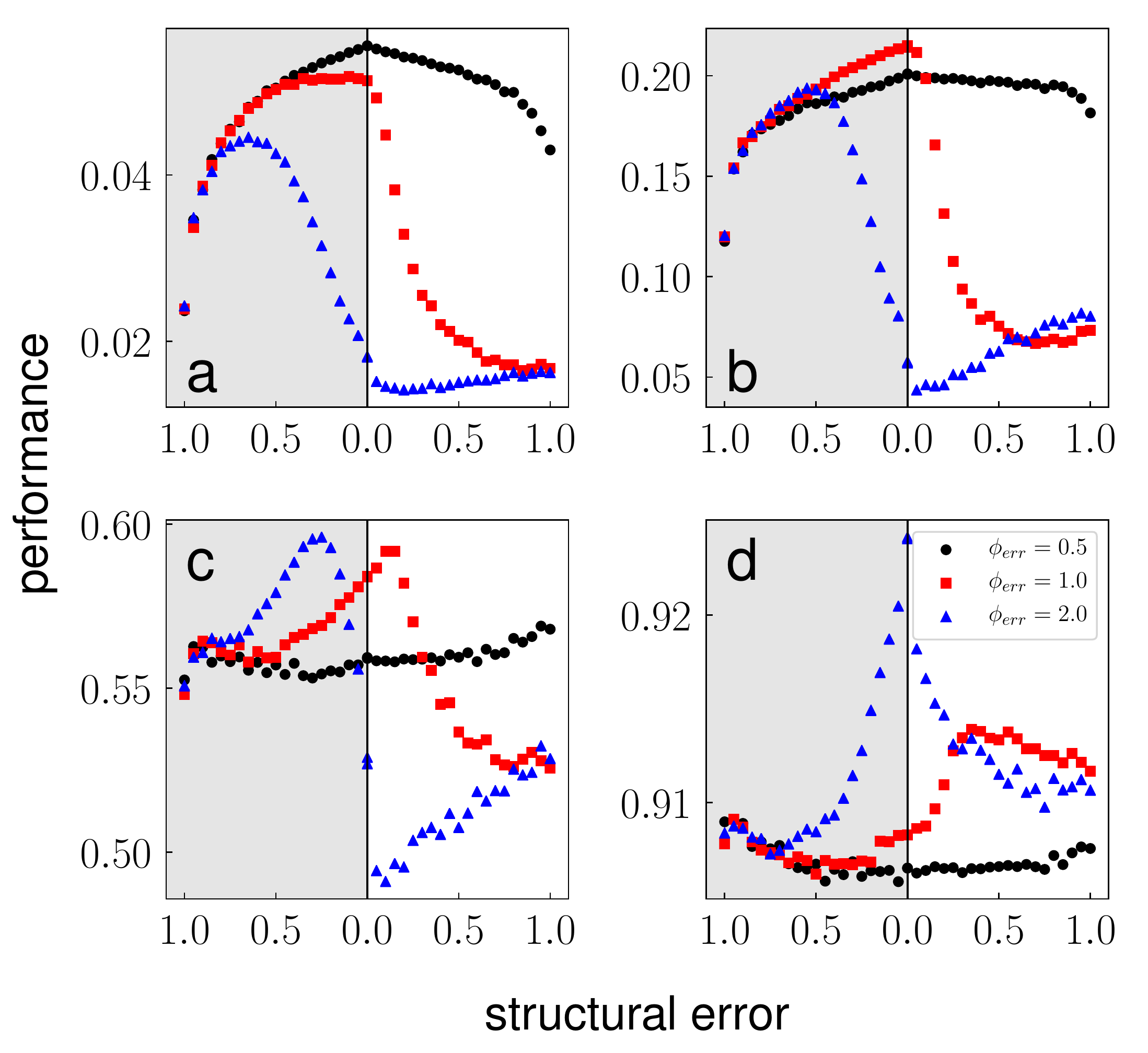}
\caption{Performance of the top spreaders on
a spatially embedded network
in presence of structural and
  dynamical noise.  Same analysis as in Fig.~\ref{fig:2}, but for a different
  network. Here, the true network structure is
given by the US power grid network of Ref.~\cite{watts1998collective}.
}
\label{fig:4}
\end{figure}

So far we reported results only for a specific network. However, our main
findings are not very sensitive to this choice, in the sense that our qualitative
results are very similar for all real networks we analyzed (see
Supplementary Material, SM).
The only major difference arises for networks characterized by strong spatial
embedding (hence, a specific modular structure identified by very loose intermodule connections \cite{Faqeeh16}). There, the strong asymmetry between the operations of
altering the network structure by adding
or removing random edges is apparent.~This fact is for instance visible in Figure~\ref{fig:4}, where we
report the analogue of Figure~\ref{fig:2} for the power grid network
originally considered in Ref.~\cite{watts1998collective}.
From the figure, we see that  $P(Q_{err})$ doesn't behave
smoothly around $\epsilon_{del} = \epsilon_{add} = 0$.
However, the
general findings valid for the two different sources of structural
noise are almost identical to those valid for networks with no
spatial embedding.

The same qualitative results hold for other values of the size of the set
of seed nodes $|Q_{err}|$, as long as its value is large enough compared to the
size of the network $N$. In the SM, we report results valid for
$|Q_{err}| = 10$ for the email network considered here in the main
text for which $N=1,133$. In such a case,
the clear pattern of Figure~\ref{fig:2} becomes
much noisier. For networks of smaller size, we observe that
the pattern is already clear even for $|Q_{err}| = 10$.

\begin{figure}[!htb]
\includegraphics[width=0.45\textwidth]{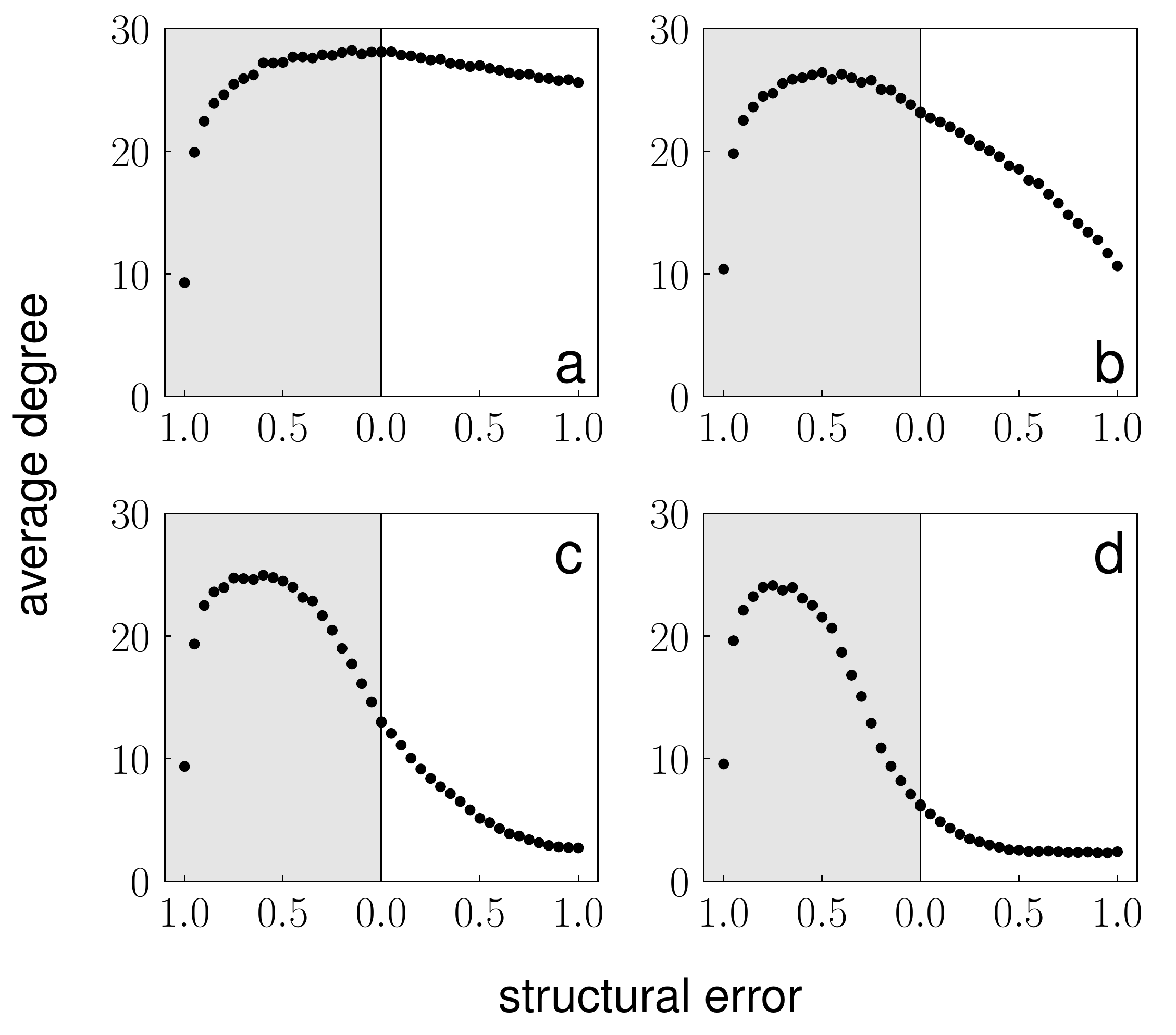}
\caption{Average degree of the set of top spreaders. We consider the same network as in Fig.~\ref{fig:1}.
(a) Average degree of the set of $|Q_{err}| = 100$ of top spreaders
identified for $\phi_{err} = 0.5$ and different values of the
structural
errors $\epsilon_{del}$ and $\epsilon_{add}$.
As in Figure~\ref{fig:2}, we use left part of the plot, highlighted
with a gray-shaded background, to report results valid for $0 \leq
\epsilon_{del} \leq 1 $
and $\epsilon_{add} =0$. The non-shaded part of the graph represents instead  results for
$0 \leq \epsilon_{add} \leq 1 $ and $\epsilon_{del} =0$.
(b, c and d) Same as in panel a, but for $\phi_{err} = 1.0$,
$\phi_{err} = 1.5$, and $\phi_{err} = 2.0$, respectively.
}
\label{fig:5}
\end{figure}

One may explain our results with the following
naive argument. For simplicity, let us
consider only the case where noise randomly
deletes true edges.  In our prior knowledge,
every true edge becomes invisible
with probability $\epsilon_{del}$.~Further,
we believe that the ICM has spreading
probability $\phi_{err}$. In
summary,
prior knowledge forces us  to think that the
effective spreading
probability on a random edge
that we are considering in the true, but unknown, network is
$(1 - \epsilon_{del}) \phi_{err}$, rather than the actual true value
$\phi_{true}$. Best predictions should be obtained for $(1 - \epsilon_{del}) \phi_{err}
\simeq \phi_{true}$.~If $\phi_{err} > \phi_{true}$, one can correct
the mistake by choosing appropriately $\epsilon_{del} \in [0, 1]$.
If $\phi_{err} < \phi_{true}$, there is no way to
satisfy the previous equation.
The best performance would be naively expected
for $\epsilon_{del} = 0$, as this value corresponds to the noise level
that minimizes the difference between effective and true
spreading probability. However, this is not what we observe in our
numerical results, where best performance is actually
achieved for $\epsilon_{del} = 1$.
The apparent paradox can be solved by accounting for structural
correlations. As it is well known,
true top spreaders in the ICM depend on the
critical regime~\cite{radicchi2017fundamental}. In the subcritical regime, central nodes
are generally better locations for seeds. In the supercritical regime
instead, peripheral nodes are selected first. In both
regimes, seeds are generally placed on nodes that are not directly
connected, as a source of spreading that is too
redundant is generally not optimal.
If the probability $\epsilon_{del}$ of random deletion of edges is not
very high, then
the ranking based on the degree centrality of the nodes is basically
unaffected. However, pairs of truly
connected nodes appear as disconnected in the noisy version of the network
regardless of their degree. As a result, for $\phi_{err} < 1$
many high-degree nodes are chosen as seeds. However, they may behave
poorly in terms of seed set, as they constitute
a source of spreading that is too redundant to be optimal.
A visual intuition of this structural explanation is provided in
Figure~\ref{fig:5}. There, we consider the true value of the
average degree of the set
of top spreaders identified using noisy information. Each panel
corresponds to a different value of $\phi_{err}$. Please note that the
value of
the average degree measured at
$\epsilon_{del} = \epsilon_{add} = 0$ is the one that corresponds
to the true set of optimal seeds for the dynamical regime $\phi_{true} = \phi_{err}$.
As expected, for the subcritical regime, the identified set of top spreaders has high
average degree and the structural noise doesn't affect much the value of this
variable,  except when $\epsilon_{del} \simeq 1$. In the supercritical regimes instead, best performance is achieved for
sets with low values of the average degree, comparable with the
average degree of the network. Structural noise changes dramatically
the set of seeds, especially in the region $0 < \epsilon_{del} <
1$. However, in
the regime of very large noise, the average degree of
the seed set is basically equal to the average degree of the network
as nodes are chosen using (almost) no structural information.

\section{Conclusions}
In this paper, we
considered a simple yet practically relevant scenario.~We
assumed that prior information used in the solution of the influence
maximization problem is affected by some noise, and we studied
how the quality of the solution found using noisy information
deteriorates as a function of the noise intensity. Our main finding is
that the quality of the solution always decreases monotonically with noise, if
structural and dynamical noise are considered independently.
However, when both sources of noise act simultaneously, one of them
can compensate the disruptive effect of the other.~In essence, noise affecting dynamical information may be suppressed
by additional noise at the structural level, or {\it vice versa}.
This fact is particularly apparent when structural noise is such
that random edges of the original network disappear with a certain
probability. As this is a plausible model of error that may affect
our knowledge of the true network structure~\cite{Newman17},
our results may be important in real-world applications. More in
general, the approach presented here may be used
to understand how incomplete and/or
 erroneous information at the level of network structure and dynamics
affects our ability to solve optimization problems in a meaningful way.

\acknowledgments

The authors thank C. Castellano for critical reading of the manuscript.
\c{S}E and FR acknowledge support from the National Science Foundation
(CMMI-1552487). AF and FR acknowledge support from the US Army Research Office
(W911NF-16-1-0104). AF acknowledges support from the Science Foundation Ireland (16/IA/4470).


\clearpage
\newpage


\section*{Supplementary material}

# Appendix: Influence maximization in noisy networks

Şirag Erkol,[1] Ali Faqeeh,[1, 2] and Filippo Radicchi[1]

[1]Center for Complex Networks and Systems Research,
School of Informatics, Computing, and Engineering,
Indiana University, Bloomington, Indiana 47408, USA
[2]MACSI, Department of Mathematics & Statistics, University of Limerick, Limerick, Ireland

## I. FLUCTUATIONS IN THE SOLUTION OF THE INFLUENCE MAXIMIZATION PROBLEM

| | | | $G = 1$ $V = 1,000$ | | | $G = 20$ $V = 50$ | | |
|---|---|---|---|---|---|---|---|---|
| $\phi$ | $\epsilon_{del}$ | $\epsilon_{add}$ | $\mu$ | $\sigma$ | $S$ | $\mu$ | $\sigma$ | $S$ |
| 0.5 | 0.0 | 0.0 | 0.094 | 0.001 | 0.765 | 0.094 | 0.001 | 0.714 |
| 0.5 | 0.5 | 0.0 | 0.092 | 0.001 | 0.747 | 0.092 | 0.001 | 0.522 |
| 0.5 | 1.0 | 0.0 | 0.065 | 0.002 | 0.089 | 0.065 | 0.002 | 0.089 |
| 0.5 | 0.0 | 0.5 | 0.093 | 0.001 | 0.748 | 0.093 | 0.001 | 0.575 |
| 0.5 | 0.0 | 1.0 | 0.093 | 0.001 | 0.699 | 0.093 | 0.001 | 0.483 |
| 1.0 | 0.0 | 0.0 | 0.183 | 0.002 | 0.621 | 0.183 | 0.002 | 0.561 |
| 1.0 | 0.5 | 0.0 | 0.179 | 0.002 | 0.748 | 0.179 | 0.002 | 0.492 |
| 1.0 | 1.0 | 0.0 | 0.139 | 0.005 | 0.089 | 0.140 | 0.006 | 0.089 |
| 1.0 | 0.0 | 0.5 | 0.181 | 0.002 | 0.492 | 0.179 | 0.002 | 0.310 |
| 1.0 | 0.0 | 1.0 | 0.162 | 0.003 | 0.301 | 0.161 | 0.003 | 0.155 |
| 2.0 | 0.0 | 0.0 | 0.450 | 0.002 | 0.615 | 0.450 | 0.002 | 0.525 |
| 2.0 | 0.5 | 0.0 | 0.422 | 0.002 | 0.645 | 0.422 | 0.002 | 0.364 |
| 2.0 | 1.0 | 0.0 | 0.427 | 0.005 | 0.089 | 0.427 | 0.005 | 0.089 |
| 2.0 | 0.0 | 0.5 | 0.441 | 0.003 | 0.765 | 0.441 | 0.003 | 0.348 |
| 2.0 | 0.0 | 1.0 | 0.437 | 0.003 | 0.782 | 0.437 | 0.003 | 0.297 |

Table S1: We study the variability of the set $Q_{err}$ of top spreaders identified by the Chen *et al.* algorithm [1] in the email contact network of Ref. [2]. We consider here sets of size $|Q_{err}| = 100$, as in the majority of the results reported in the main text. For simplicity, we consider the case of absence of mistakes in prior knowledge about the dynamics of the spreading process, so that $\phi_{err} = \phi_{true} = \phi$. The value of the parameter $\phi$ is reported in the first column of the table. Values of the parameters $\epsilon_{del}$ and $\epsilon_{add}$ of the noise affecting prior knowledge of the network structure are reported in the second and third columns, respectively. First, we quantify the magnitude of finite-size fluctuations associated with the algorithm by Chen *et al.* [1]. We generate a single instance of the structural noise, i.e, $G = 1$. We then apply the identification algorithm by Chen *et al.* $V = 1,000$ times. Every time that the algorithm is applied, it relies on $R = 1,000$ realizations of the bond percolation process with occupation probability $p = \phi\, p_c$, where $p_c$ is the critical value of the spreading probability. For each set of top spreaders determined by the algorithm by Chen *et al.*, we quantify the average value of the outbreak size $\mu = O/N$ and the associated standard deviation $\sigma$ obtained in $T = 10$ simulations of the spreading dynamics. The numerical values reported in the fourth and fifth columns are further averaged over the $V = 1,000$ applications of the identification algorithm. We quantify also the self-consistency $S$ of the $V$ sets found [3]. Specifically, $S = \sum_i z_i^2 / \sum_i z_i$, where $z_i$ is the fraction of times that node $i$ is identified in the set $Q_{err}$ of top spreaders. As a reference, consider that $S = 1$ indicates that the set is always the same across the various realizations; instead, $S = |Q_{err}|/N \simeq 0.088$ indicates that nodes have equal probability to be in the set of top spreaders. Values of $S$ are reported in the sixth column. Finally, we quantify fluctuations induced by both structural noise and the identification algorithm simultaneously. We basically repeat the same analysis just described, but for $G = 20$ different realizations of the structural noise. Values of $\mu$, $\sigma$ and $S$ for this second set of experiments are reported in columns $7 - 9$. Please note that, for each realization of the structural noise, we identify the top influential spreaders only $V = 50$ times.

## II. ANALYSIS OF OTHER NETWORKS

We report here the results obtained for the analysis of the real-word networks listed in Table S2. For every network, we report two sets of figures, obtained setting $|Q_{err}| = 100$ and $|Q_{err}| = 10$. Description of the figures is identical to those appearing in the main text.

| # | Network | Size | Figures |
|---|---------|------|---------|
| 1 | URV email [2] | 1133 | Fig. S1 |
| 2 | US Air Transportation [4] | 500 | Fig. S2, Fig. S3 |
| 3 | Tennis [5] | 4342 | Fig. S4, Fig. S5 |
| 4 | C. Elegans, neural [6] | 297 | Fig. S6, Fig. S7 |
| 5 | High school, 2012 [7] | 180 | Fig. S8, Fig. S9 |
| 6 | Air traffic [8] | 1226 | Fig. S10, Fig. S11 |
| 7 | Open flights [8, 9] | 2939 | Fig. S12, Fig. S13 |
| 8 | UC Irvine [8, 10] | 1899 | Fig. S14, Fig. S15 |
| 9 | Petster, hamster [8] | 1858 | Fig. S16, Fig. S17 |
| 10 | Political blogs [11] | 1224 | Fig. S18, Fig. S19 |
| 11 | Political books [11] | 105 | Fig. S20, Fig. S21 |
| 12 | US Power grid [6] | 4941 | Fig. S22, Fig. S23 |
| 13 | S 838 [12] | 512 | Fig. S24, Fig. S25 |
| 14 | Yeast, protein [13] | 2284 | Fig. S26, Fig. S27 |

Table S2: Real-world networks analyzed in this paper. From left to right, we report: order of appearance of the network, name of the network and the reference(s) in which the network was first analyzed, number of nodes, figure numbers for the results of the analysis performed in this paper.

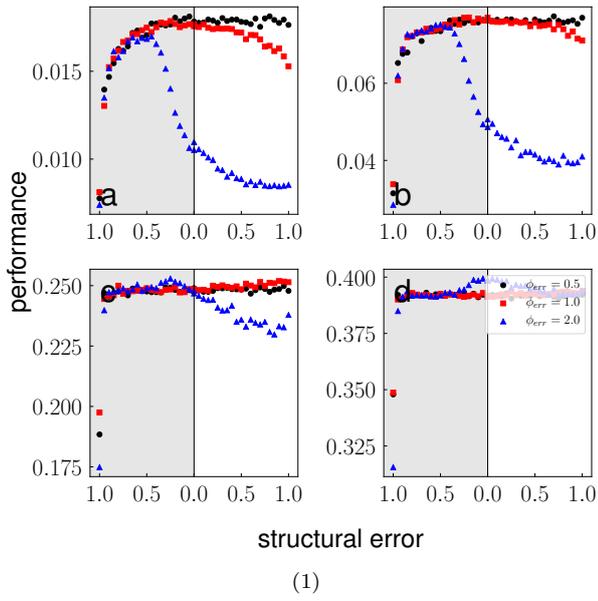

(1)

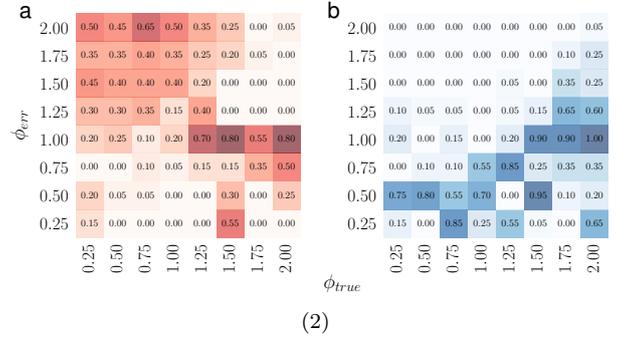

(2)

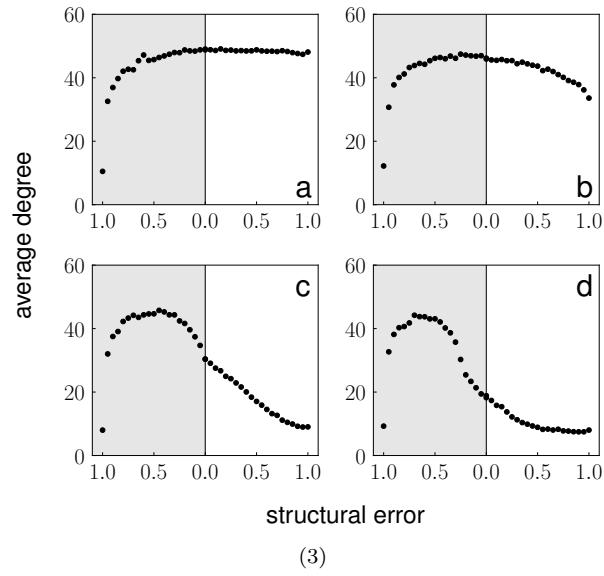

(3)

Figure S1: URV email. $|Q_{err}| = 10$.

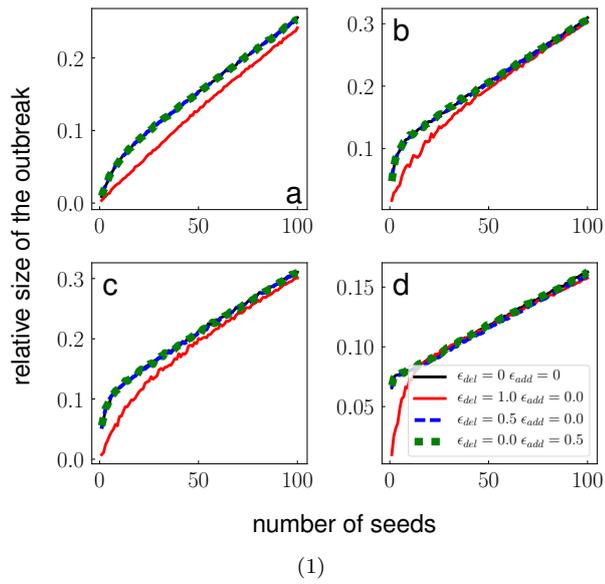

(1)

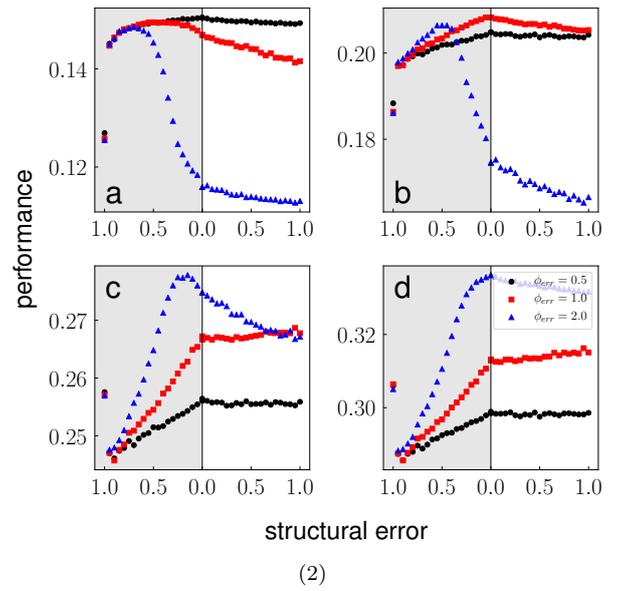

(2)

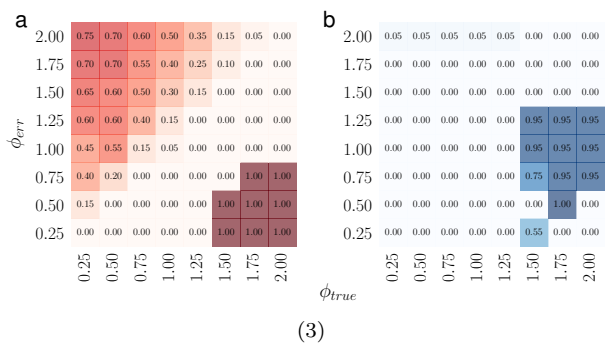

(3)

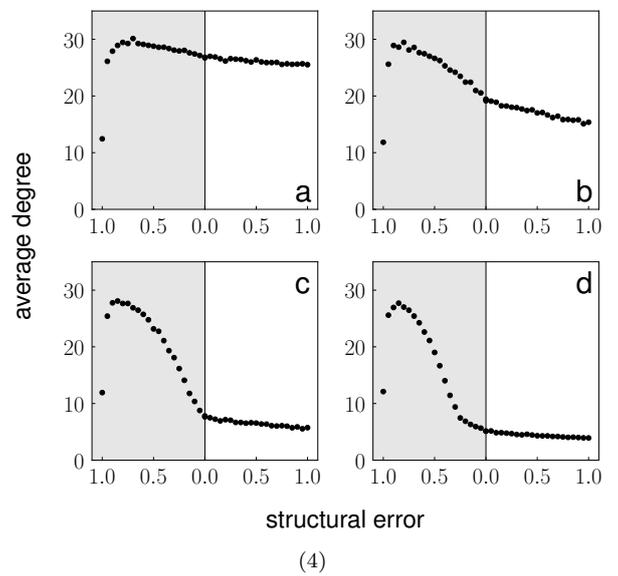

(4)

Figure S2: US Air Transportation. $|Q_{err}| = 100$

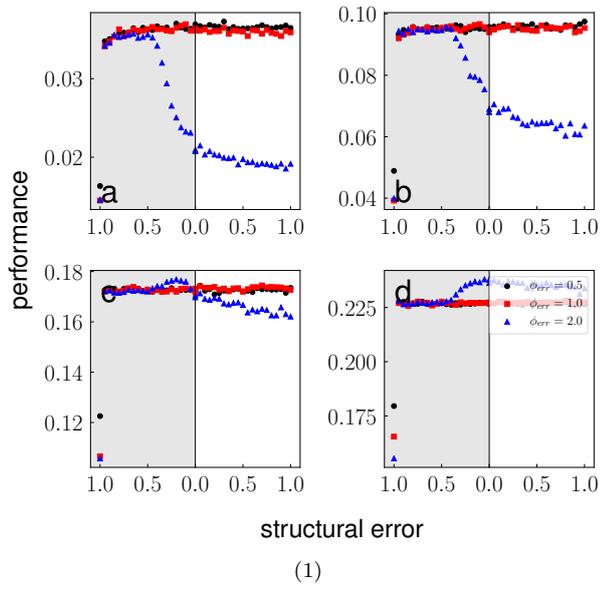

(1)

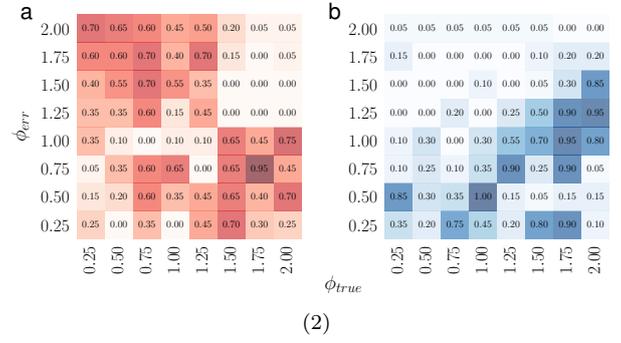

(2)

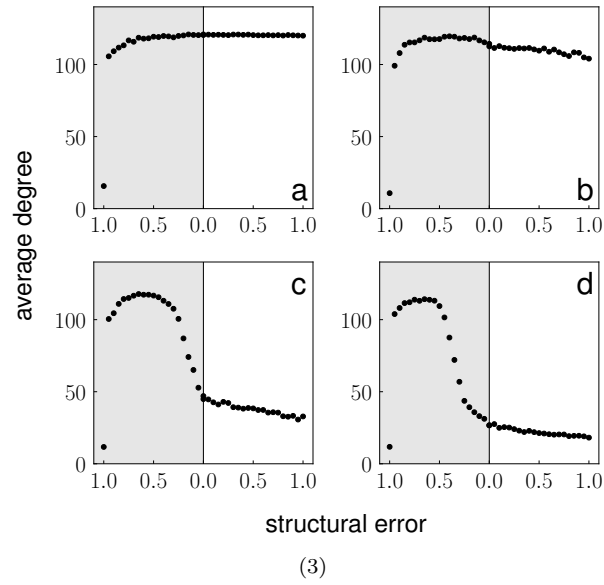

(3)

Figure S3: US Air Transportation. $|Q_{err}| = 10$

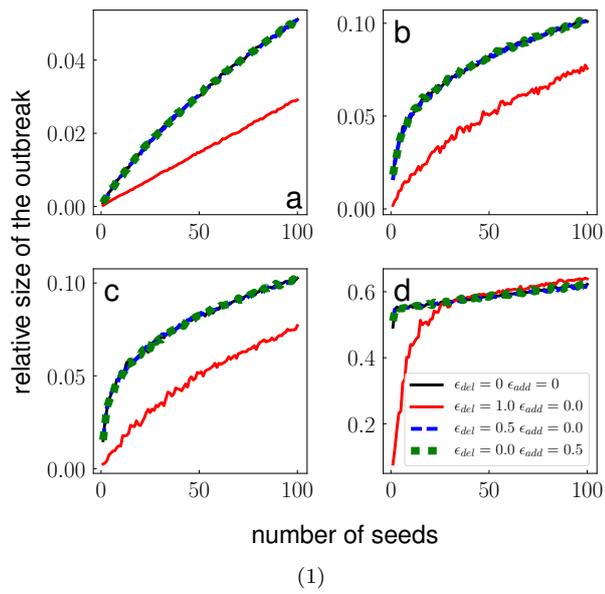

(1)

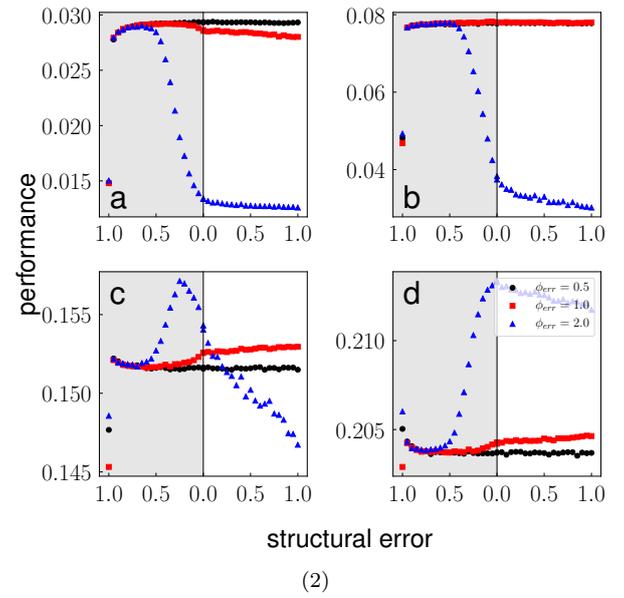

(2)

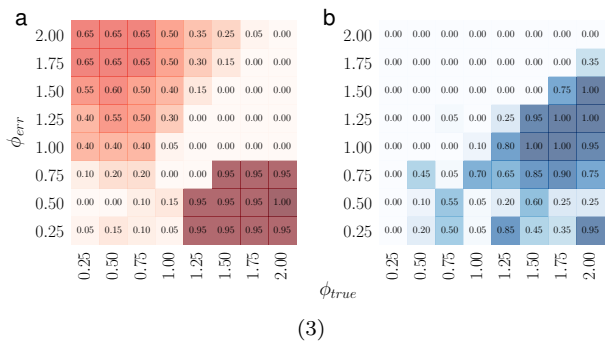

(3)

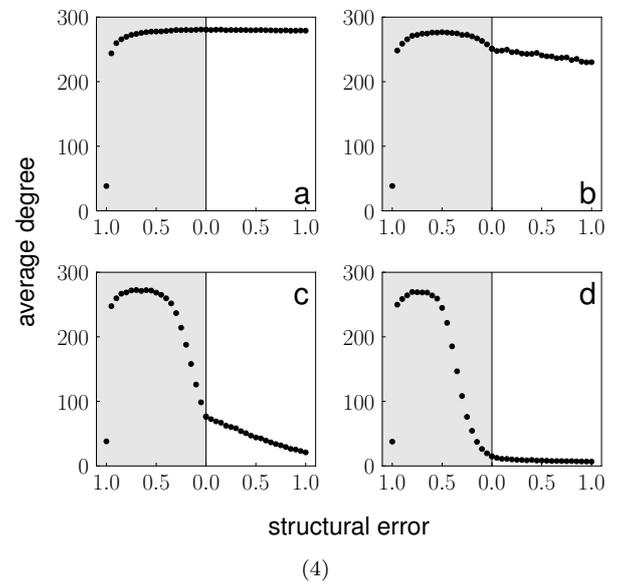

(4)

Figure S4: Tennis. $|Q_{err}| = 100$

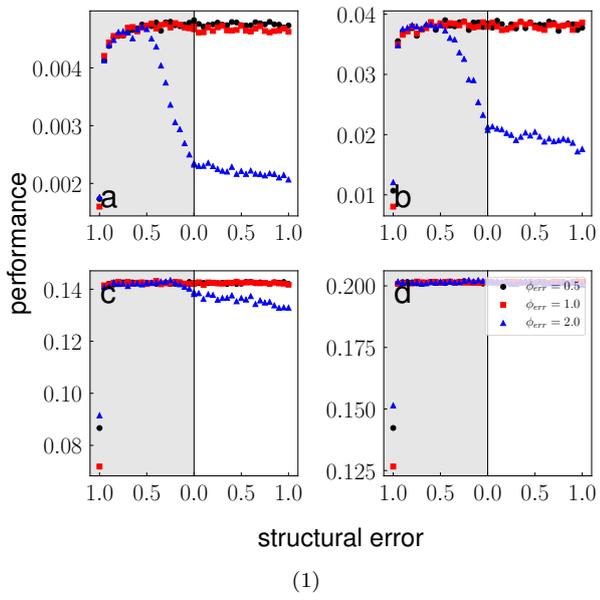

(1)

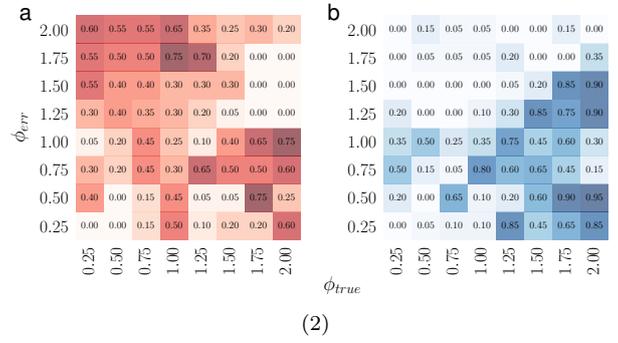

(2)

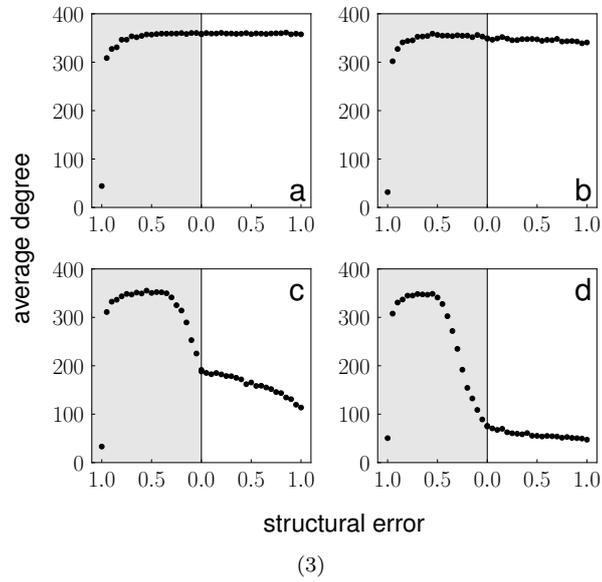

(3)

Figure S5: Tennis. $|Q_{err}| = 10$

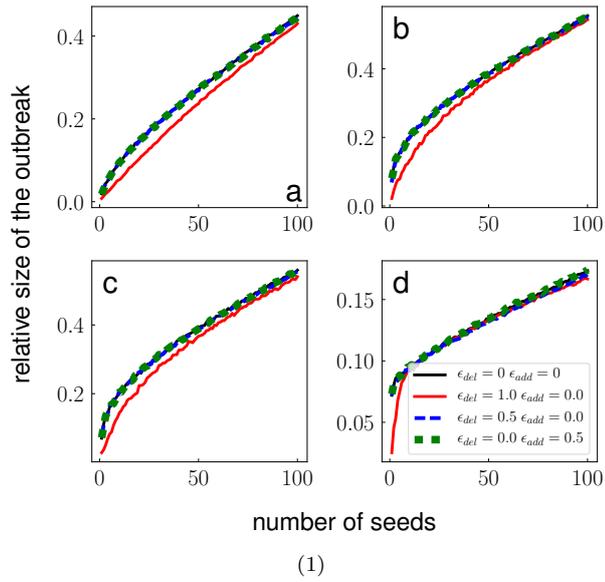

(1)

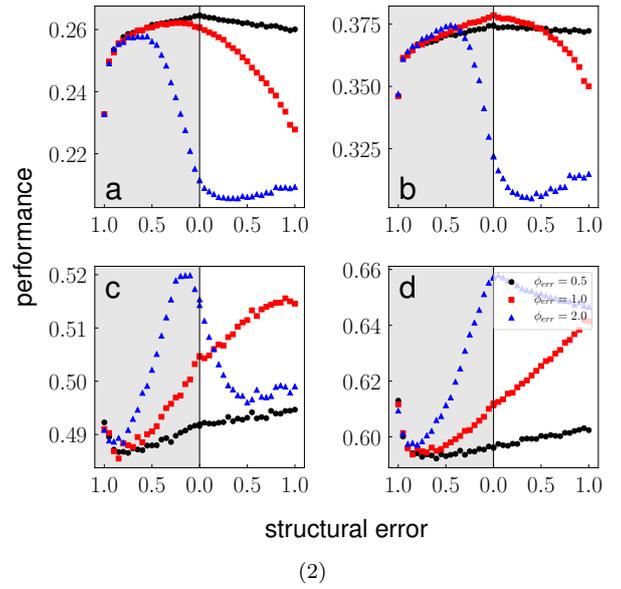

(2)

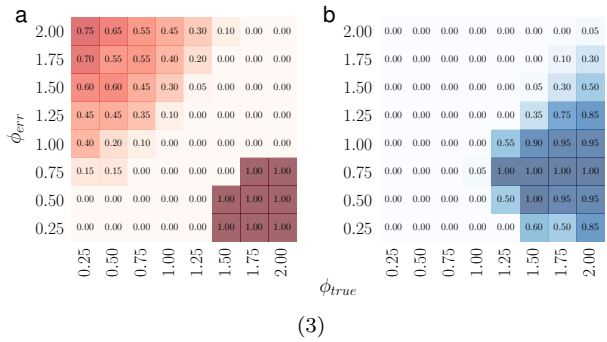

(3)

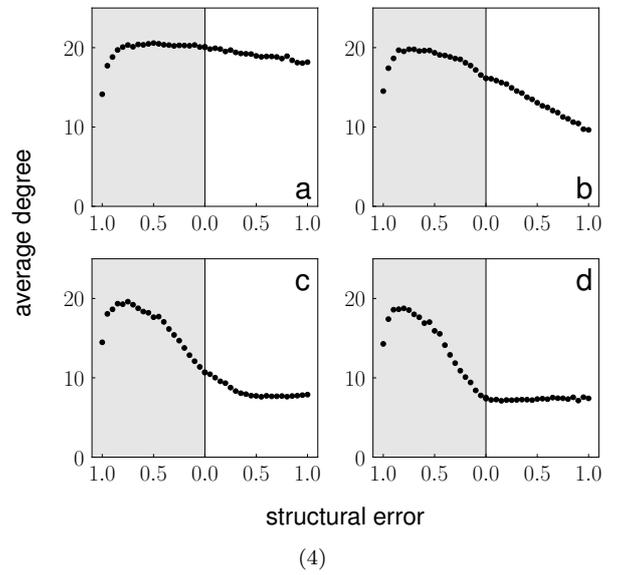

(4)

Figure S6: C. Elegans, neural. $|Q_{err}| = 100$

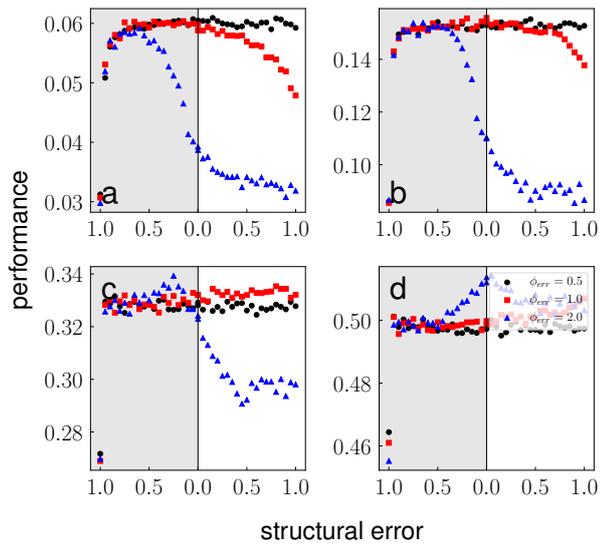

(1)

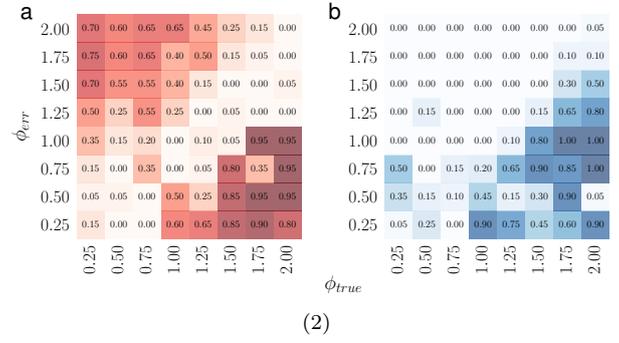

(2)

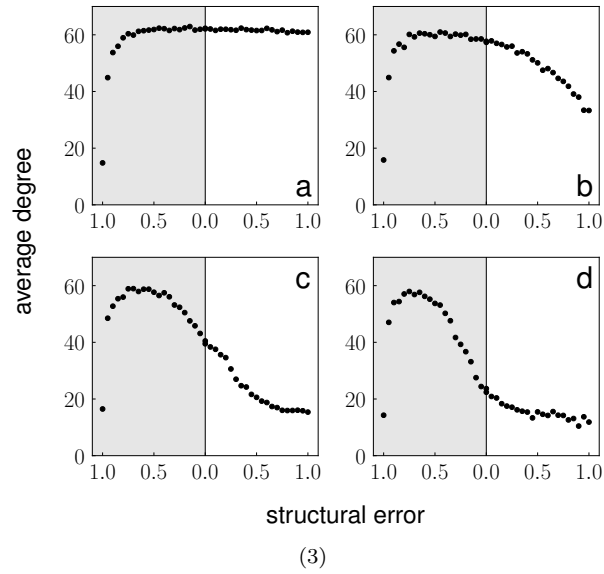

(3)

Figure S7: C. Elegans, neural. $|Q_{err}| = 10$

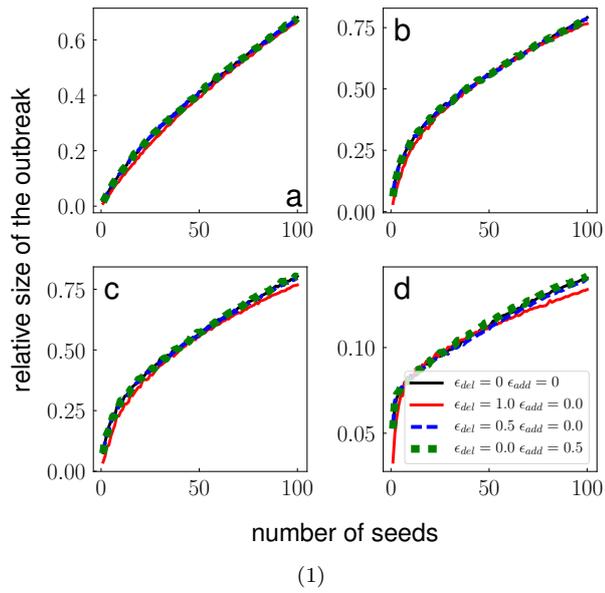

(1)

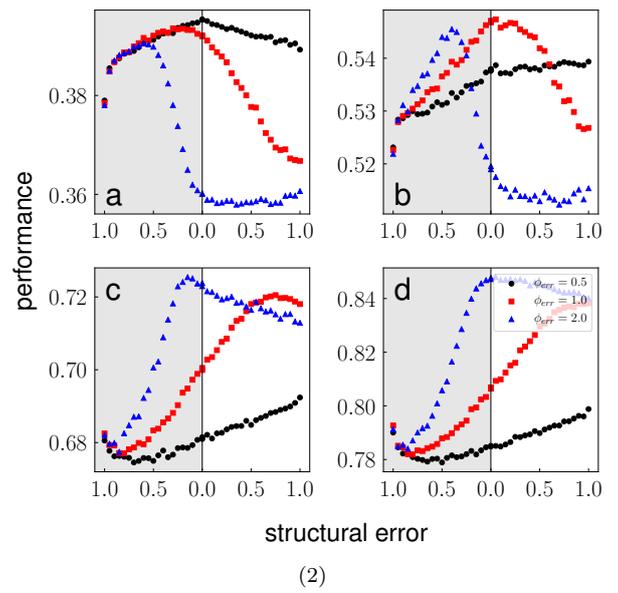

(2)

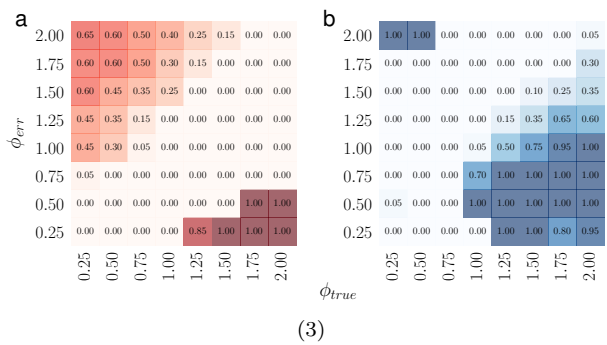

(3)

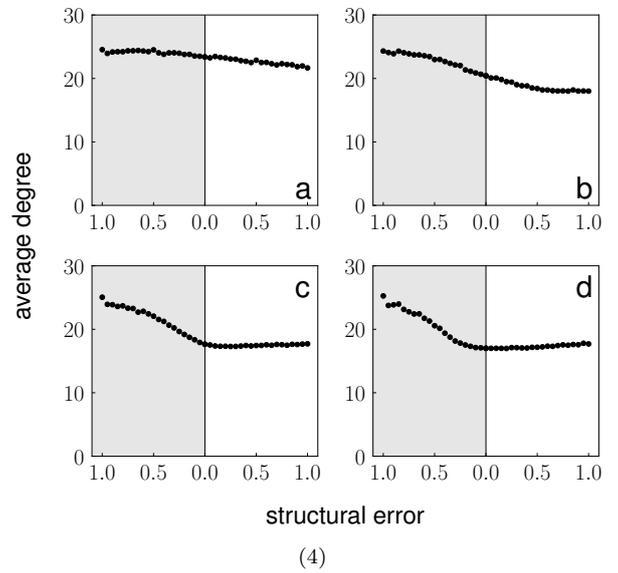

(4)

Figure S8: High school, 2012. $|Q_{err}| = 100$

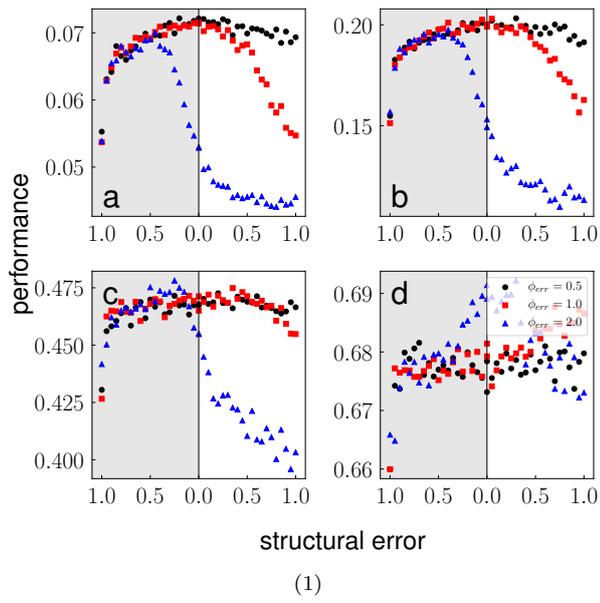

(1)

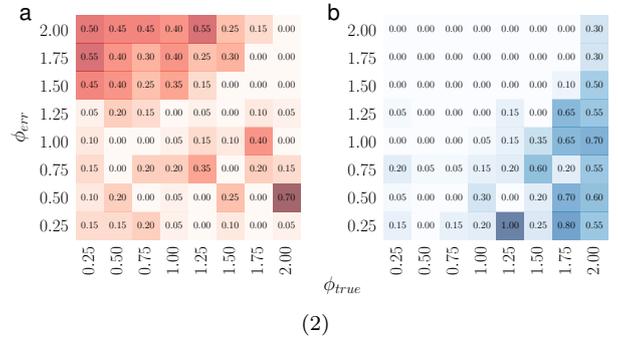

(2)

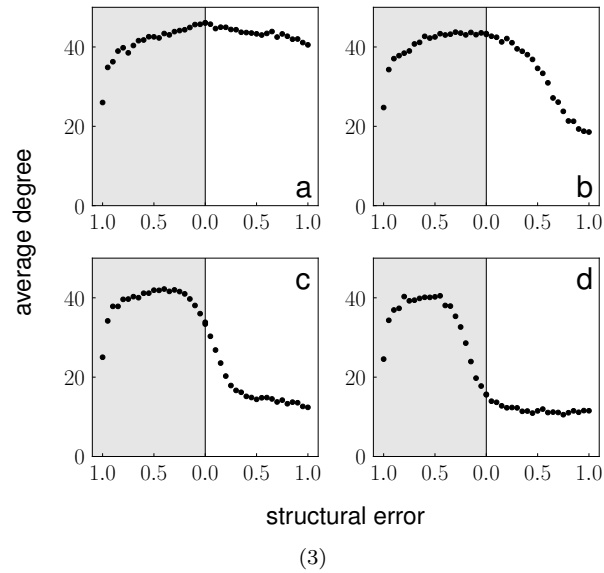

(3)

Figure S9: High school, 2012. $|Q_{err}| = 10$

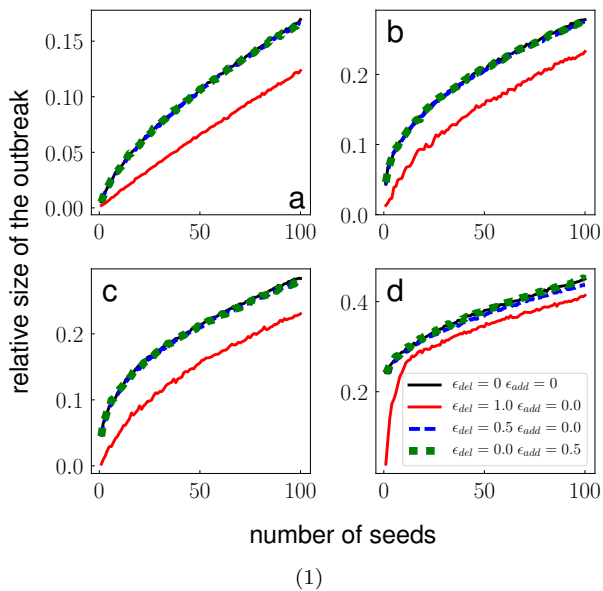

(1)

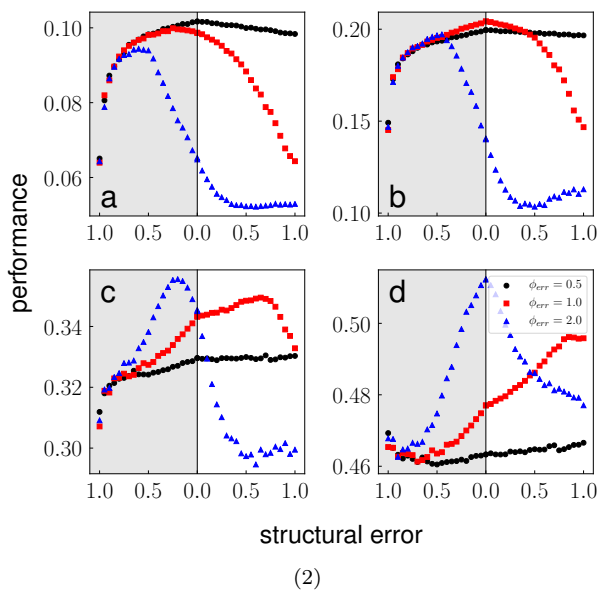

(2)

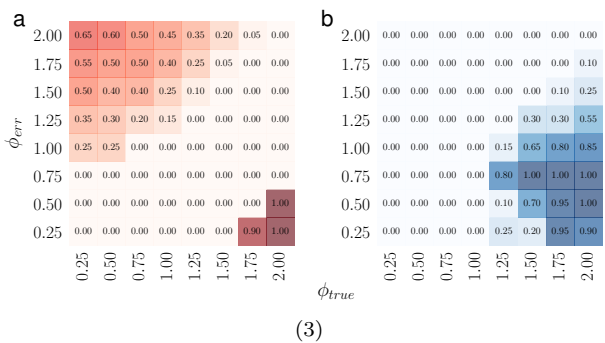

(3)

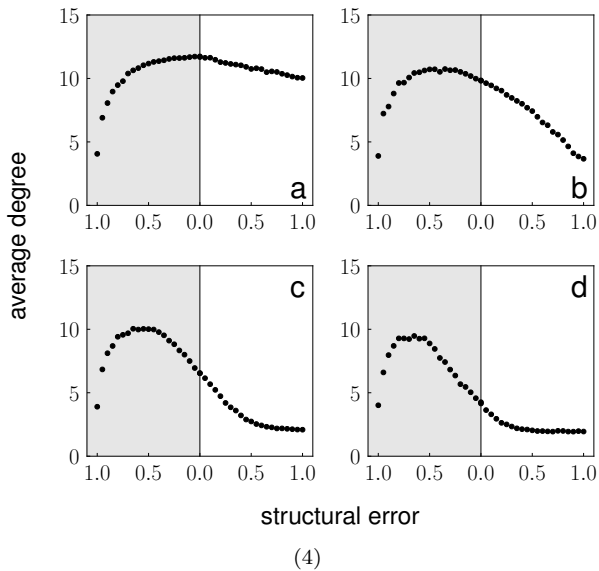

(4)

Figure S10: Air traffic. $|Q_{err}| = 100$

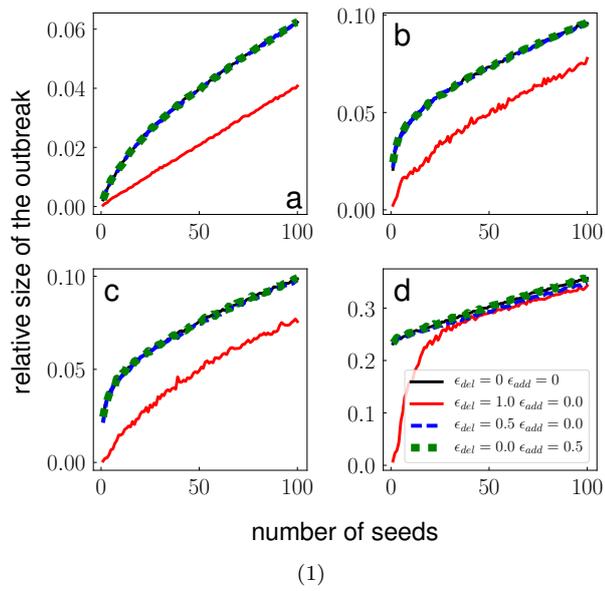

(1)

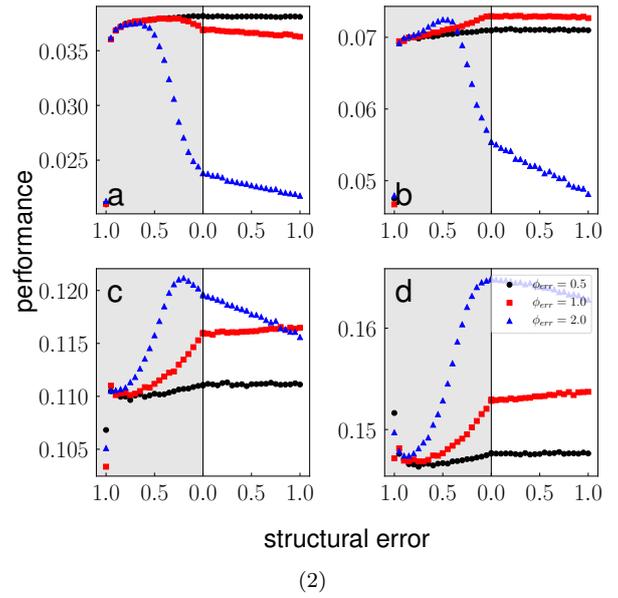

(2)

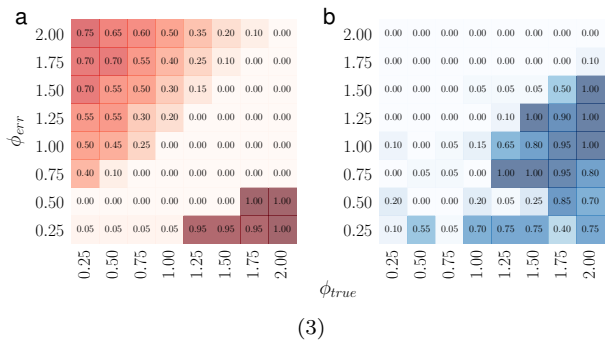

(3)

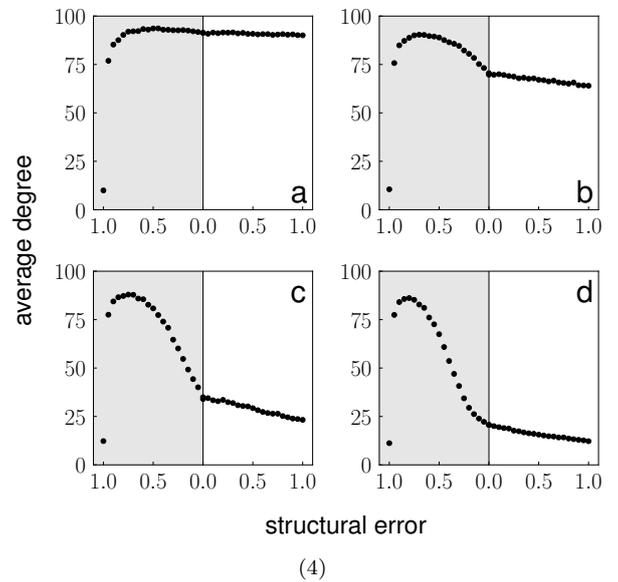

(4)

Figure S12: Open flights. $|Q_{err}| = 100$

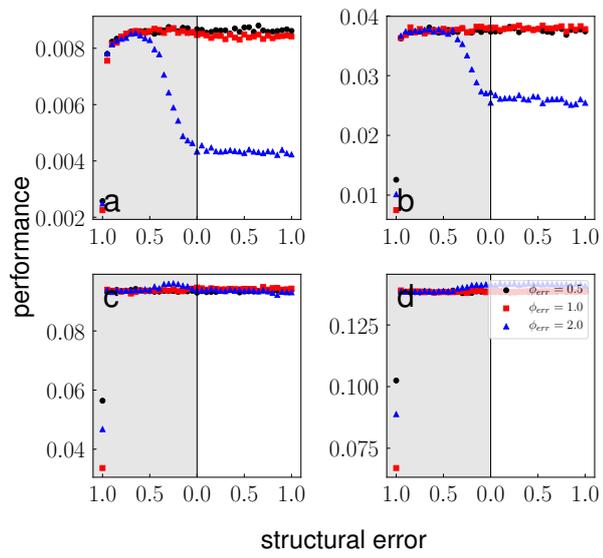

(1)

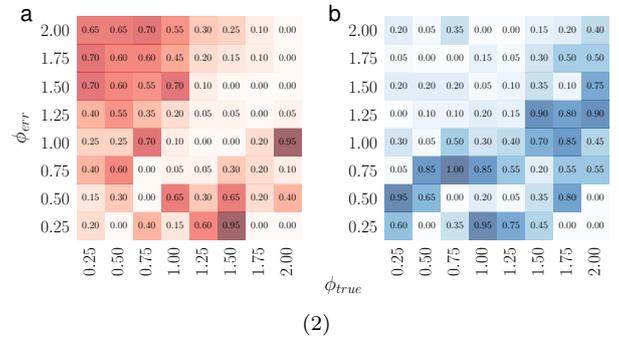

(2)

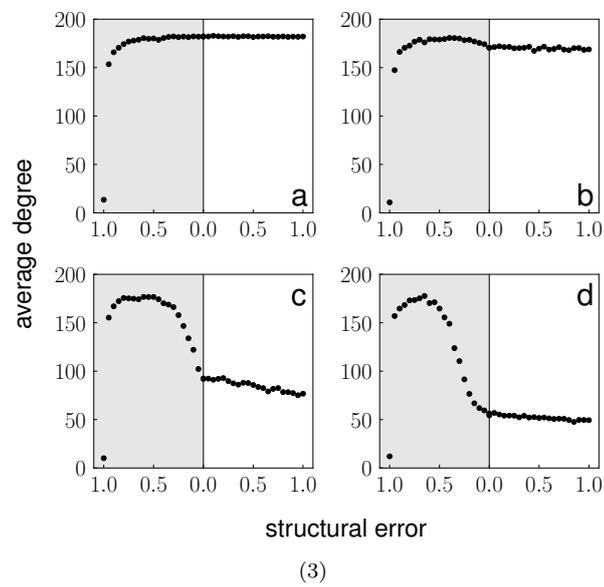

(3)

Figure S13: Open flights. $|Q_{err}| = 10$

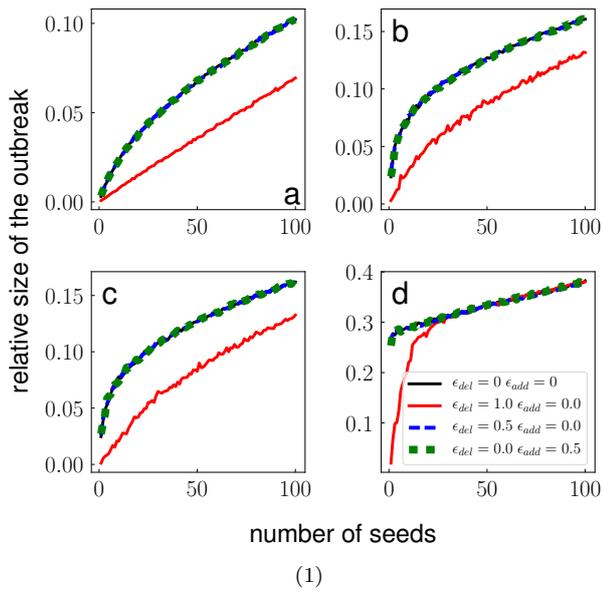

(1)

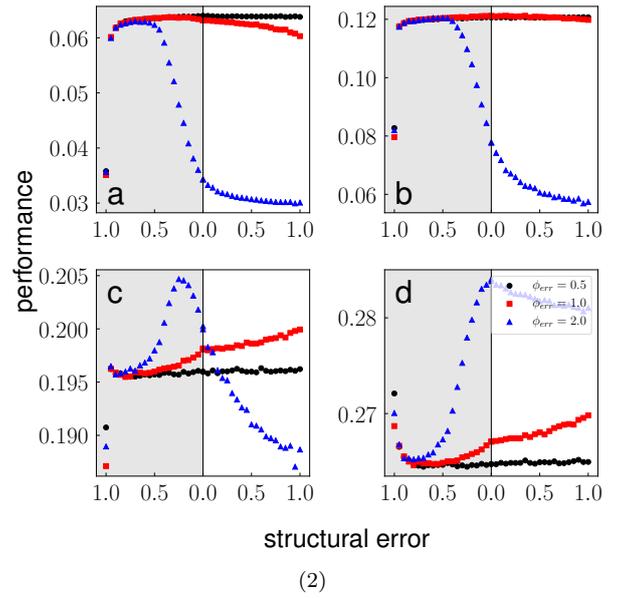

(2)

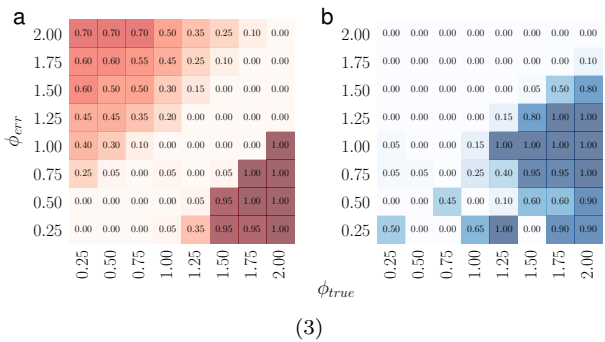

(3)

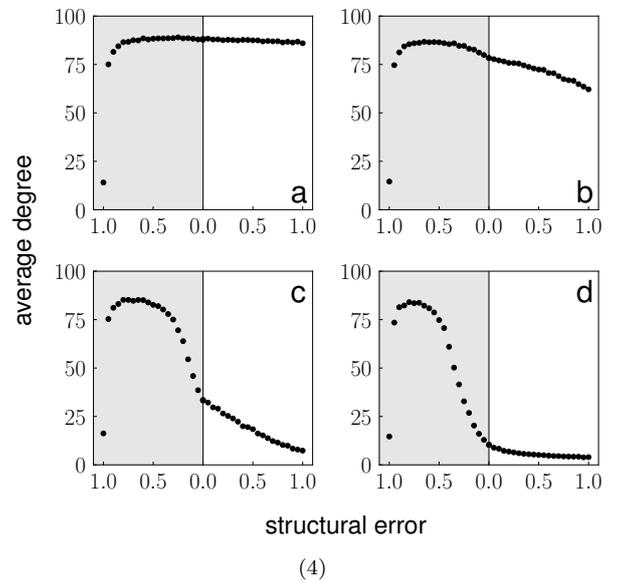

(4)

Figure S14: UC Irvine. $|Q_{err}| = 100$

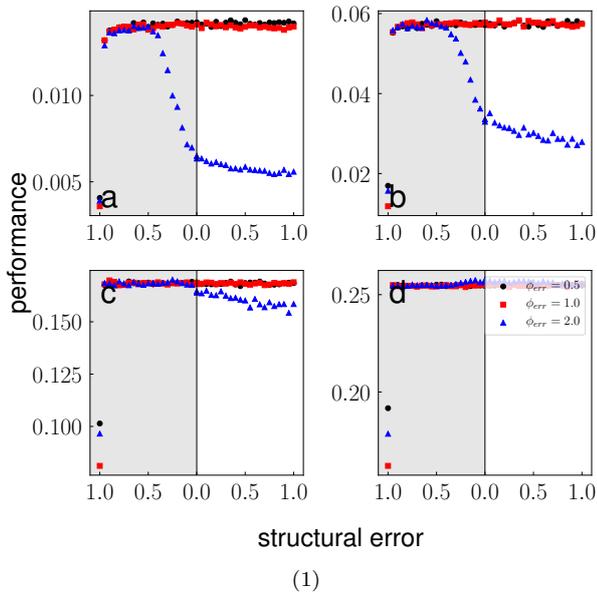

(1)

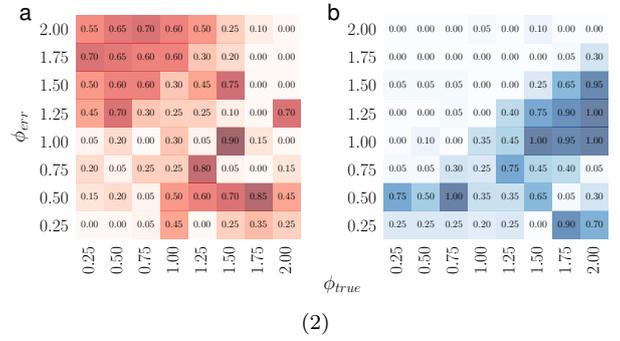

(2)

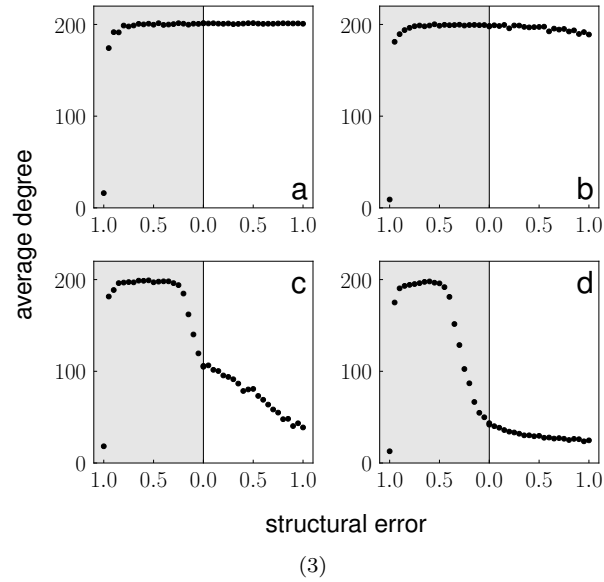

(3)

Figure S15: UC Irvine. $|Q_{err}| = 10$

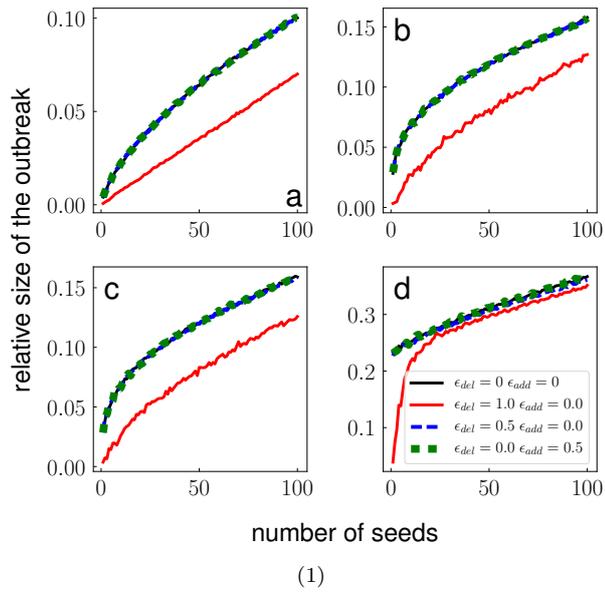

(1)

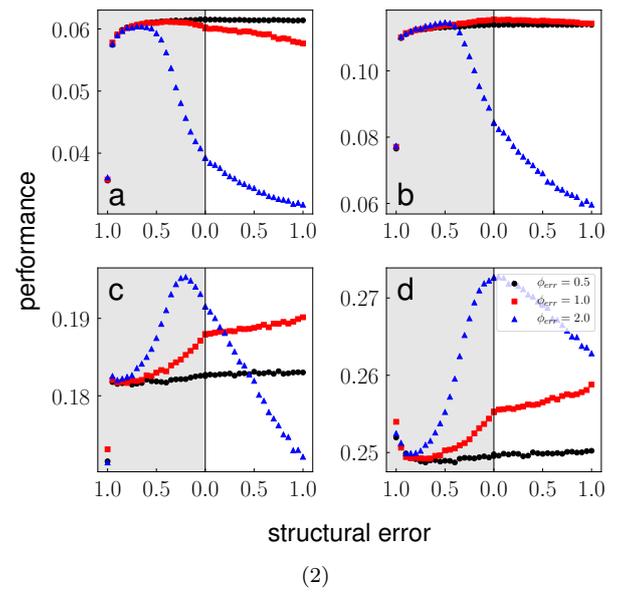

(2)

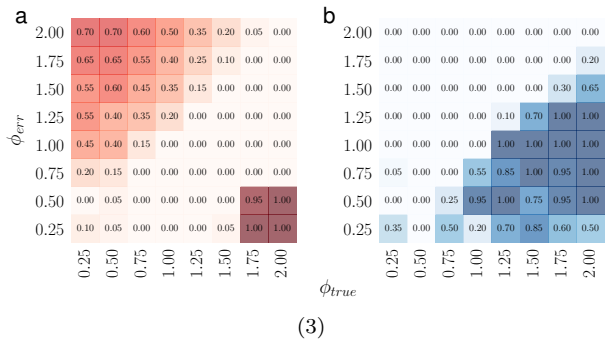

(3)

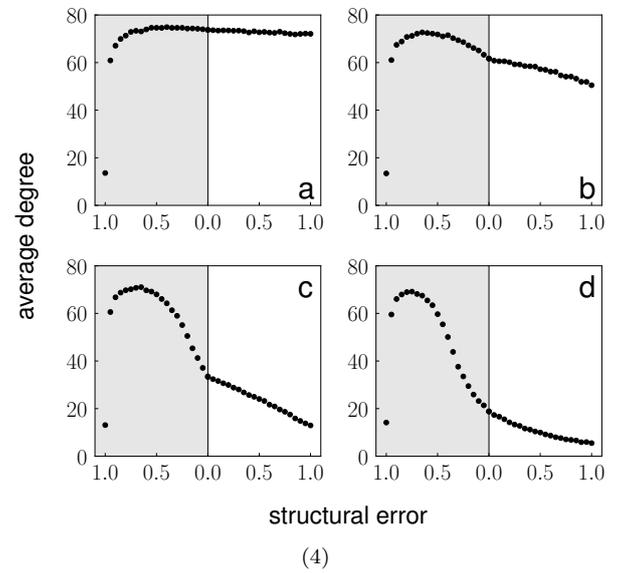

(4)

Figure S16: Petster, hamster. $|Q_{err}| = 100$

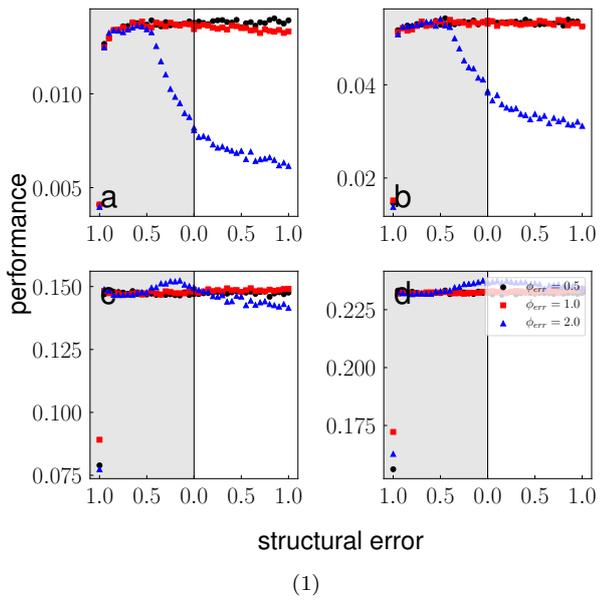

(1)

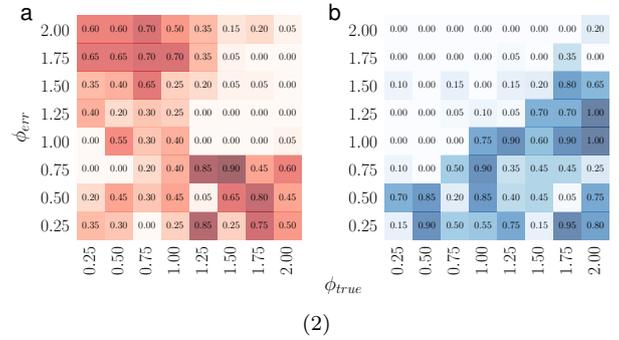

(2)

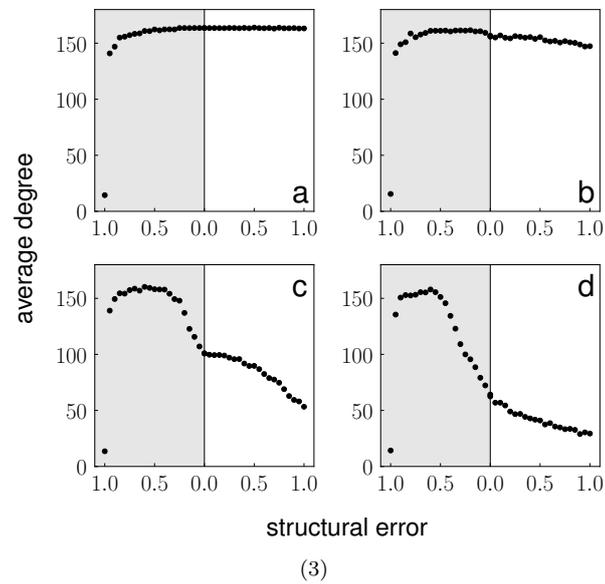

(3)

Figure S17: Petster, hamster. $|Q_{err}| = 10$

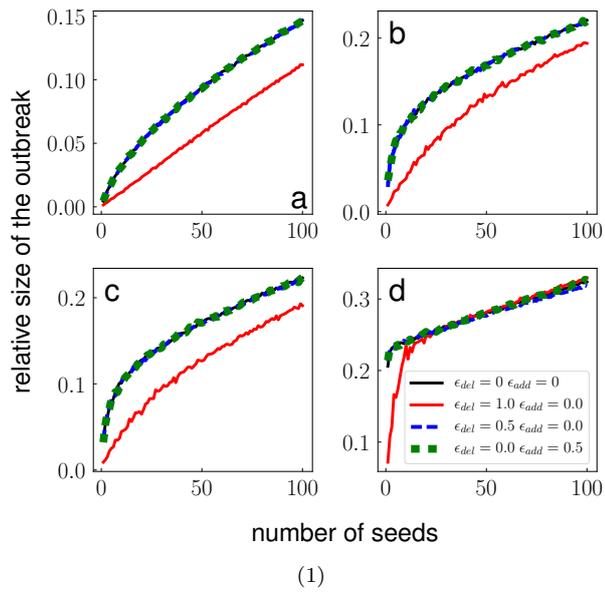

(1)

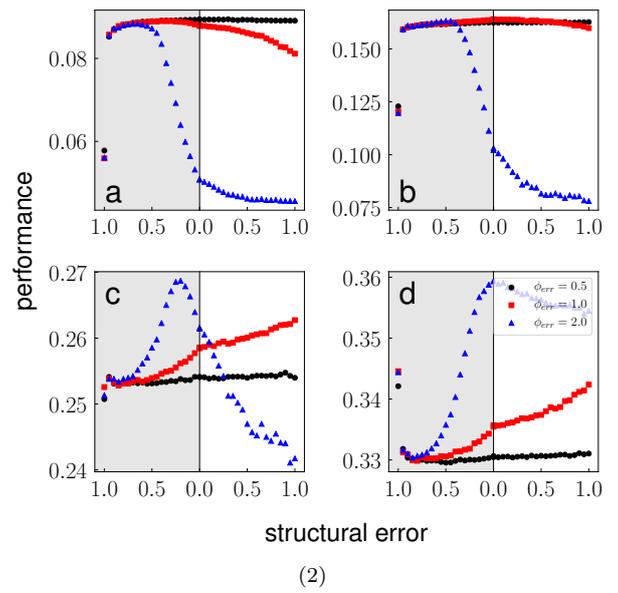

(2)

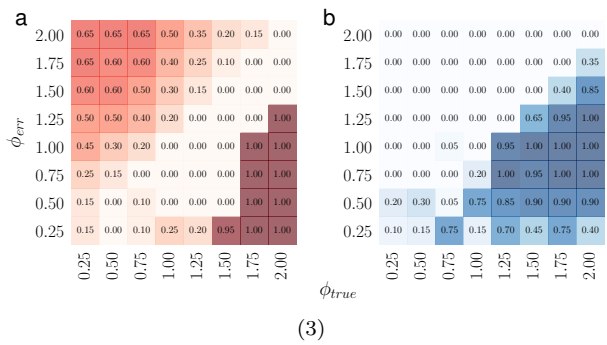

(3)

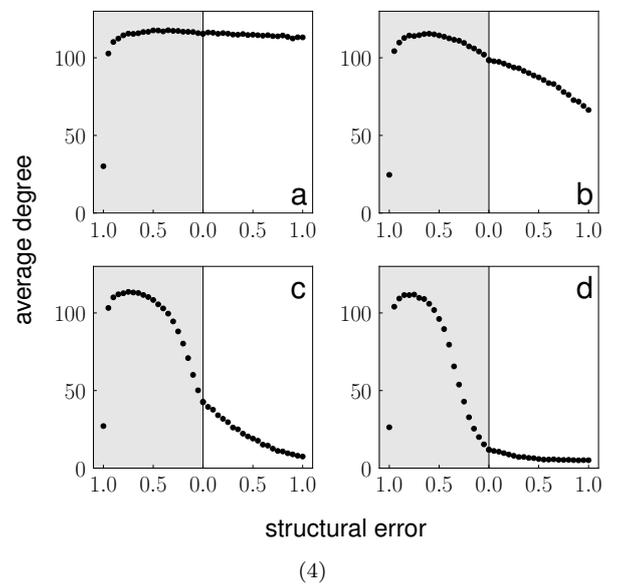

(4)

Figure S18: Political blogs. $|Q_{err}| = 100$

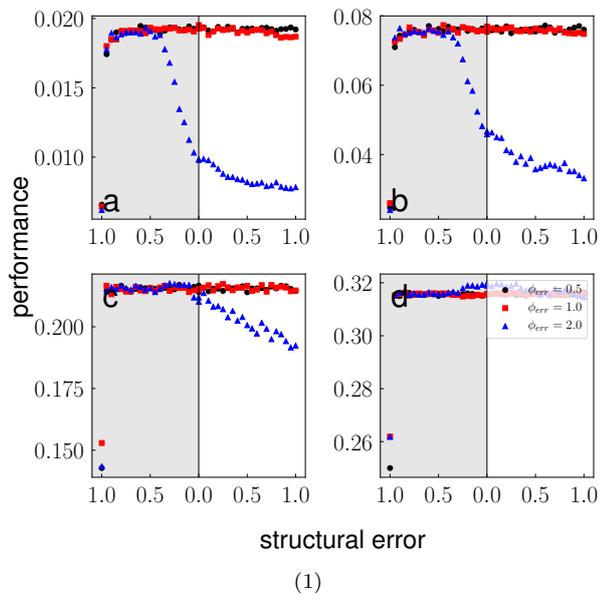

(1)

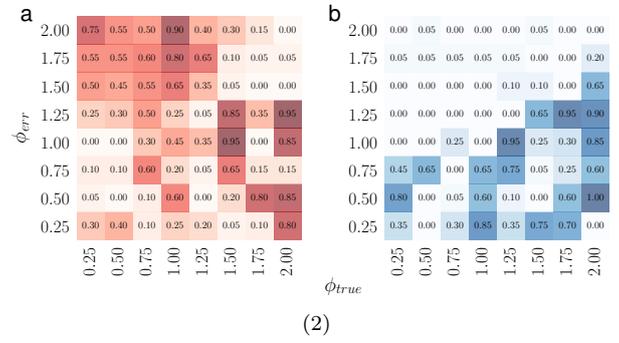

(2)

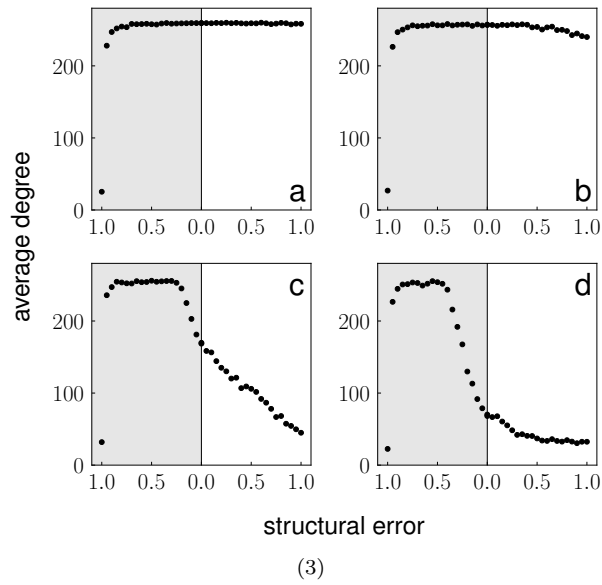

(3)

Figure S19: Political blogs. $|Q_{err}| = 10$

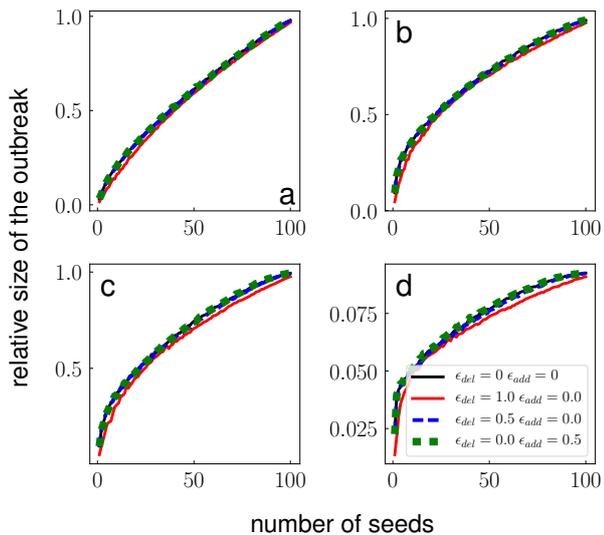

(1)

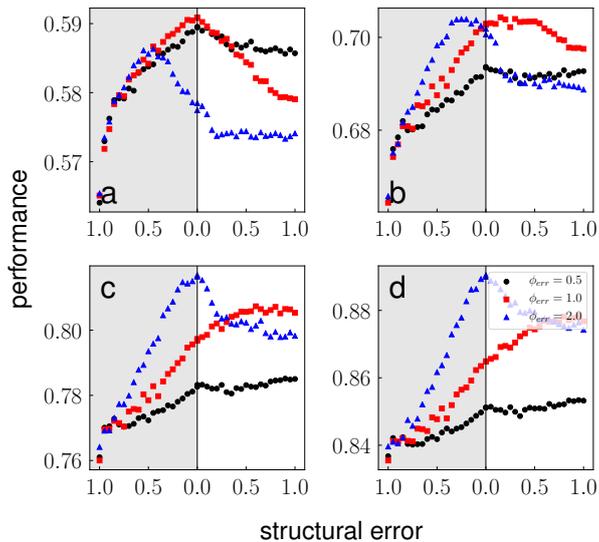

(2)

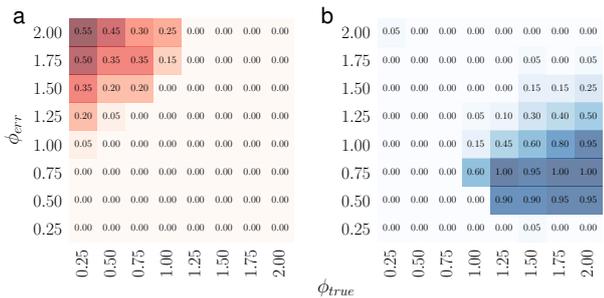

(3)

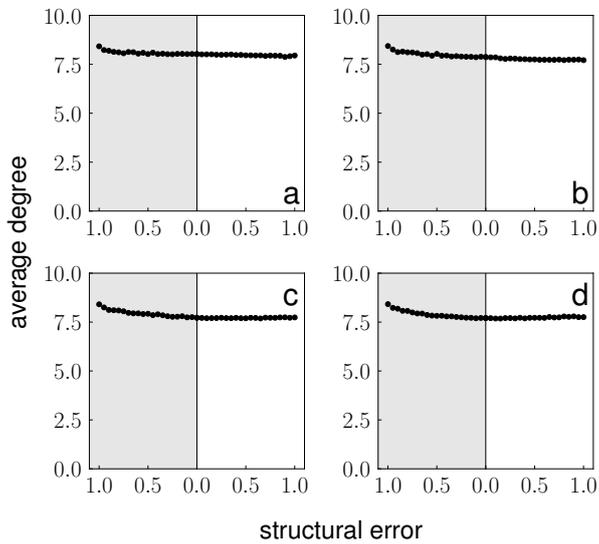

(4)

Figure S20: Political books. $|Q_{err}| = 100$

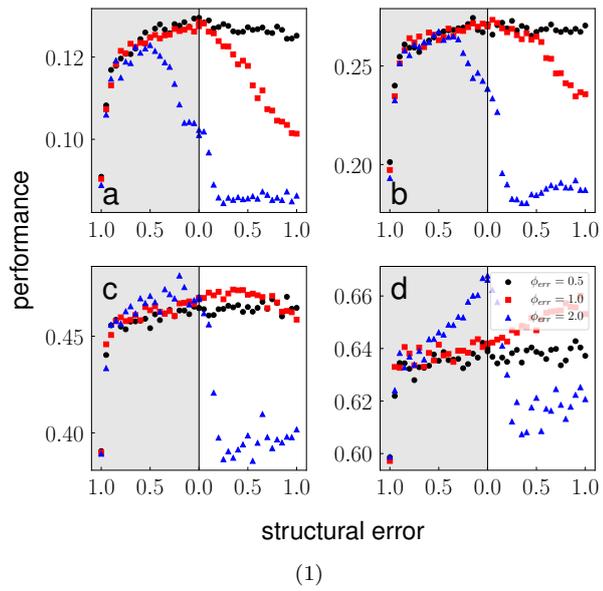

(1)

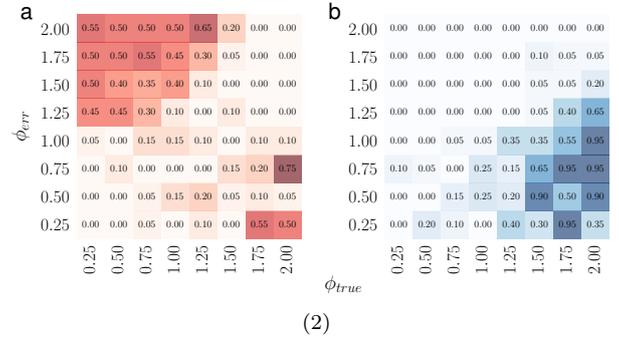

(2)

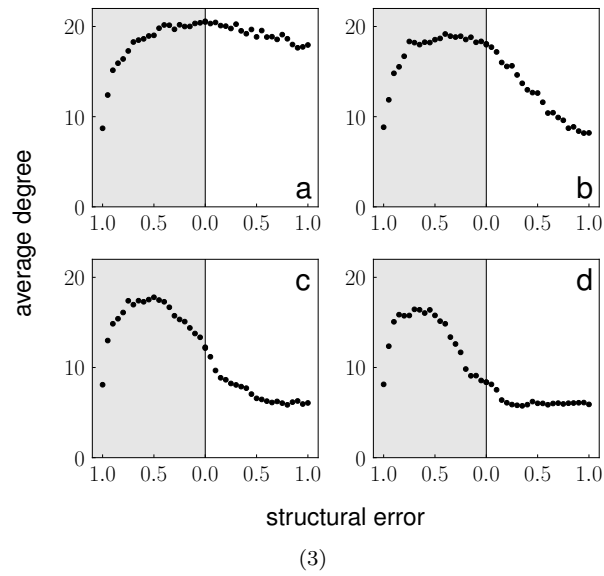

(3)

Figure S21: Political books. $|Q_{err}| = 10$

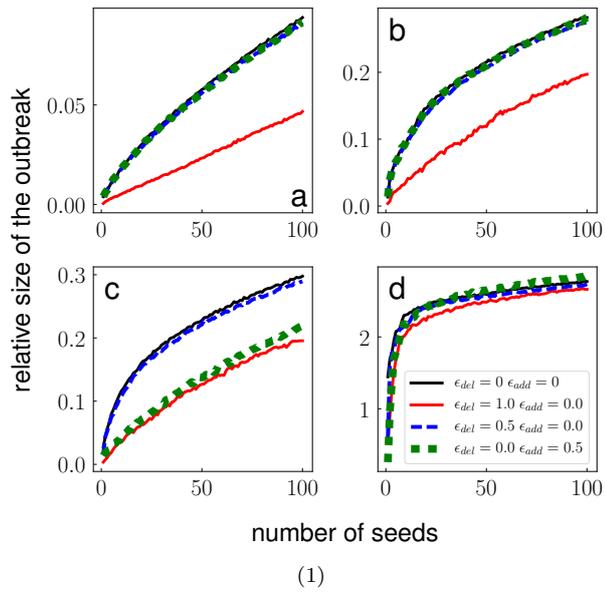

(1)

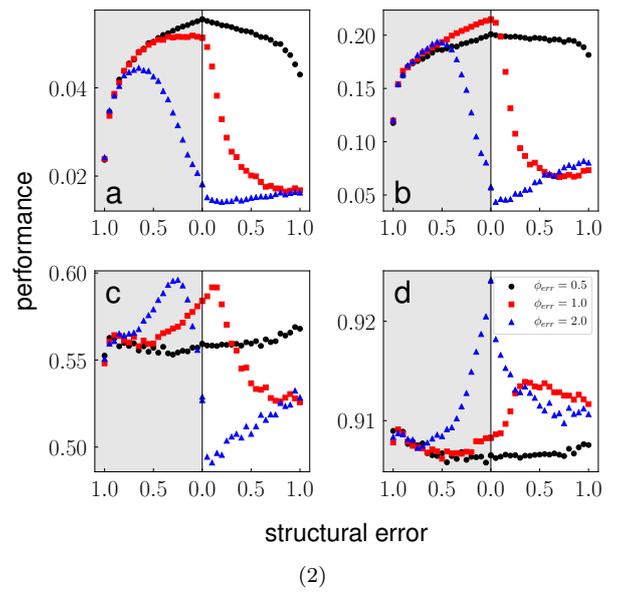

(2)

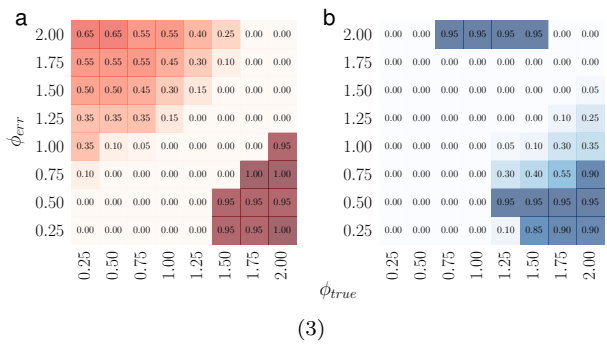

(3)

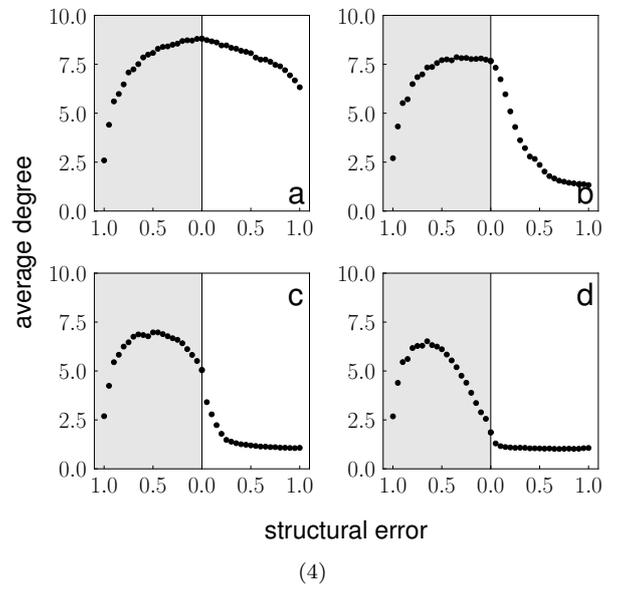

(4)

Figure S22: US power grid. $|Q_{err}| = 100$

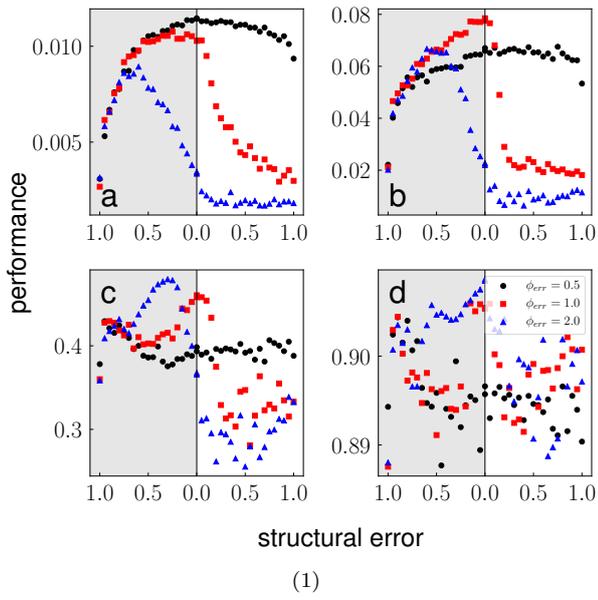

(1)

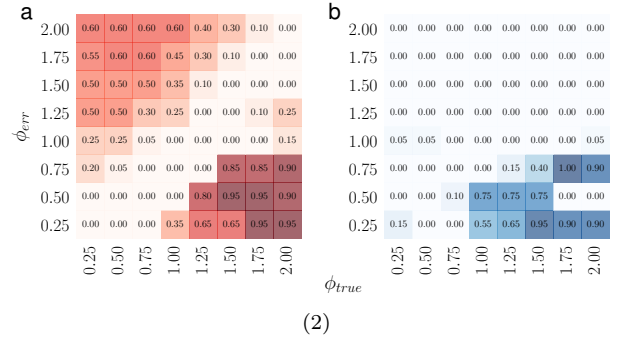

(2)

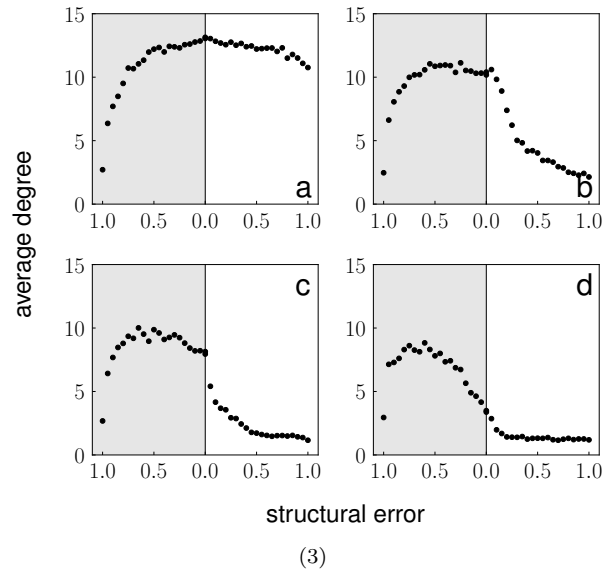

(3)

Figure S23: US power grid. $|Q_{err}| = 10$

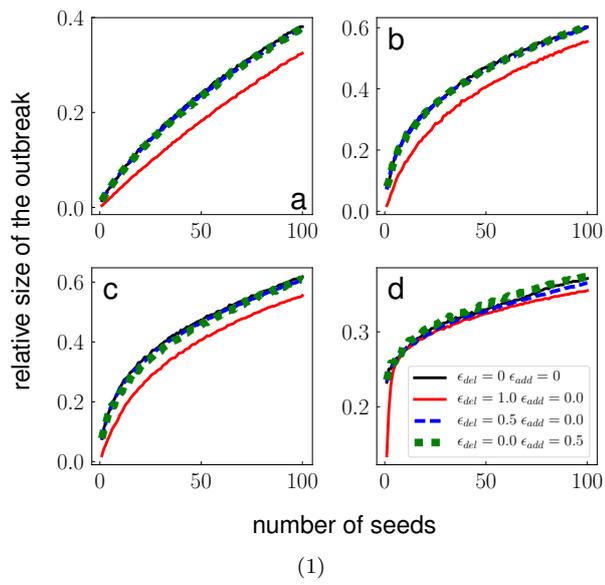

(1)

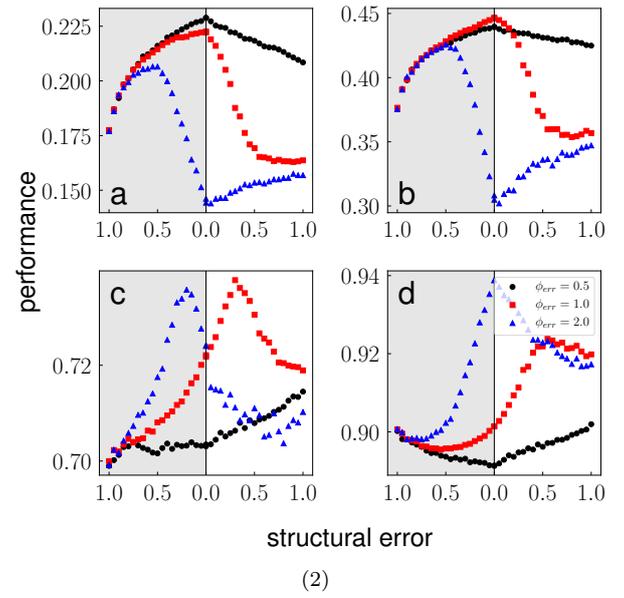

(2)

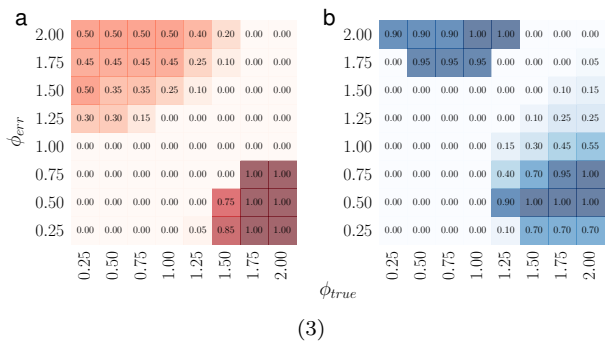

(3)

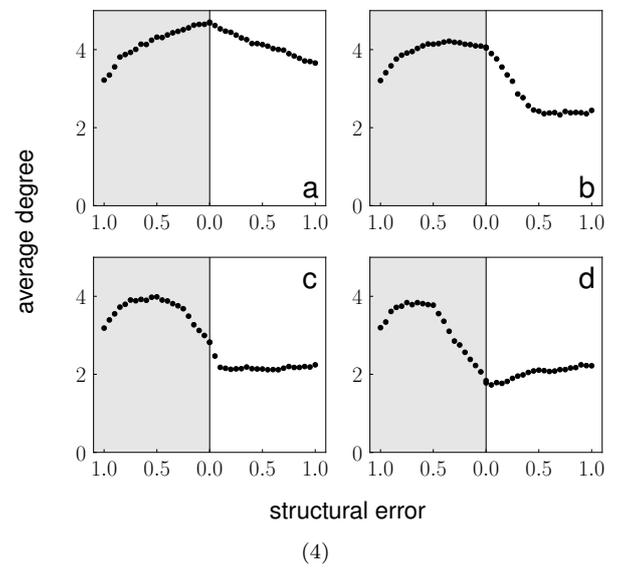

(4)

Figure S24: s838. $|Q_{err}| = 100$

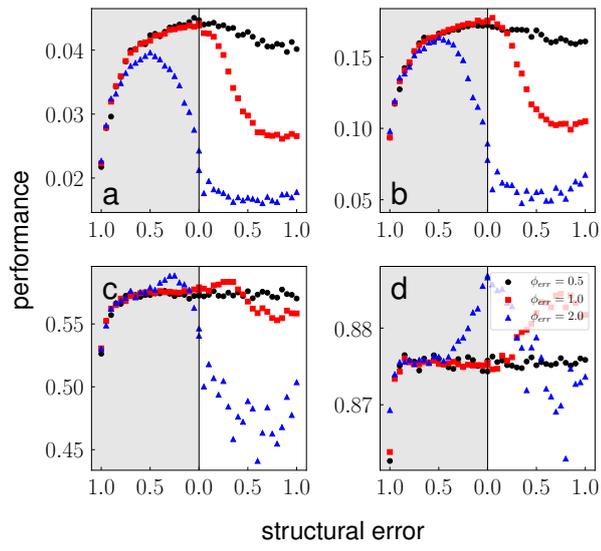

(1)

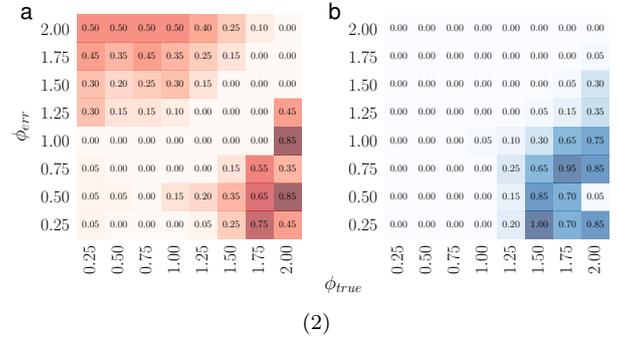

(2)

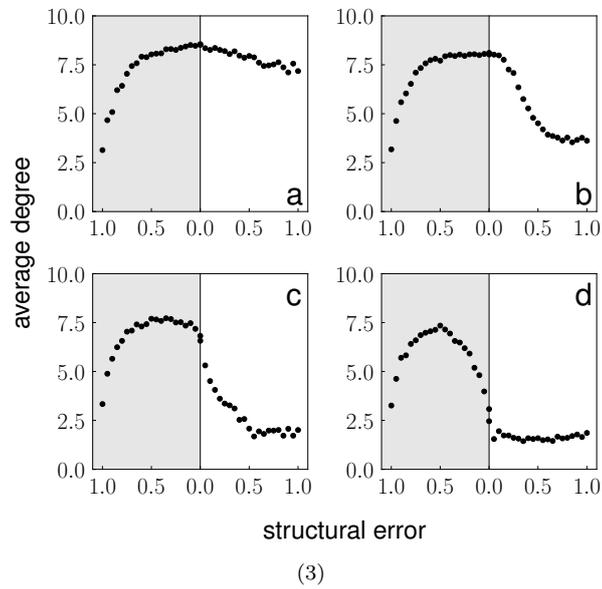

(3)

Figure S25: s838. $|Q_{err}| = 10$

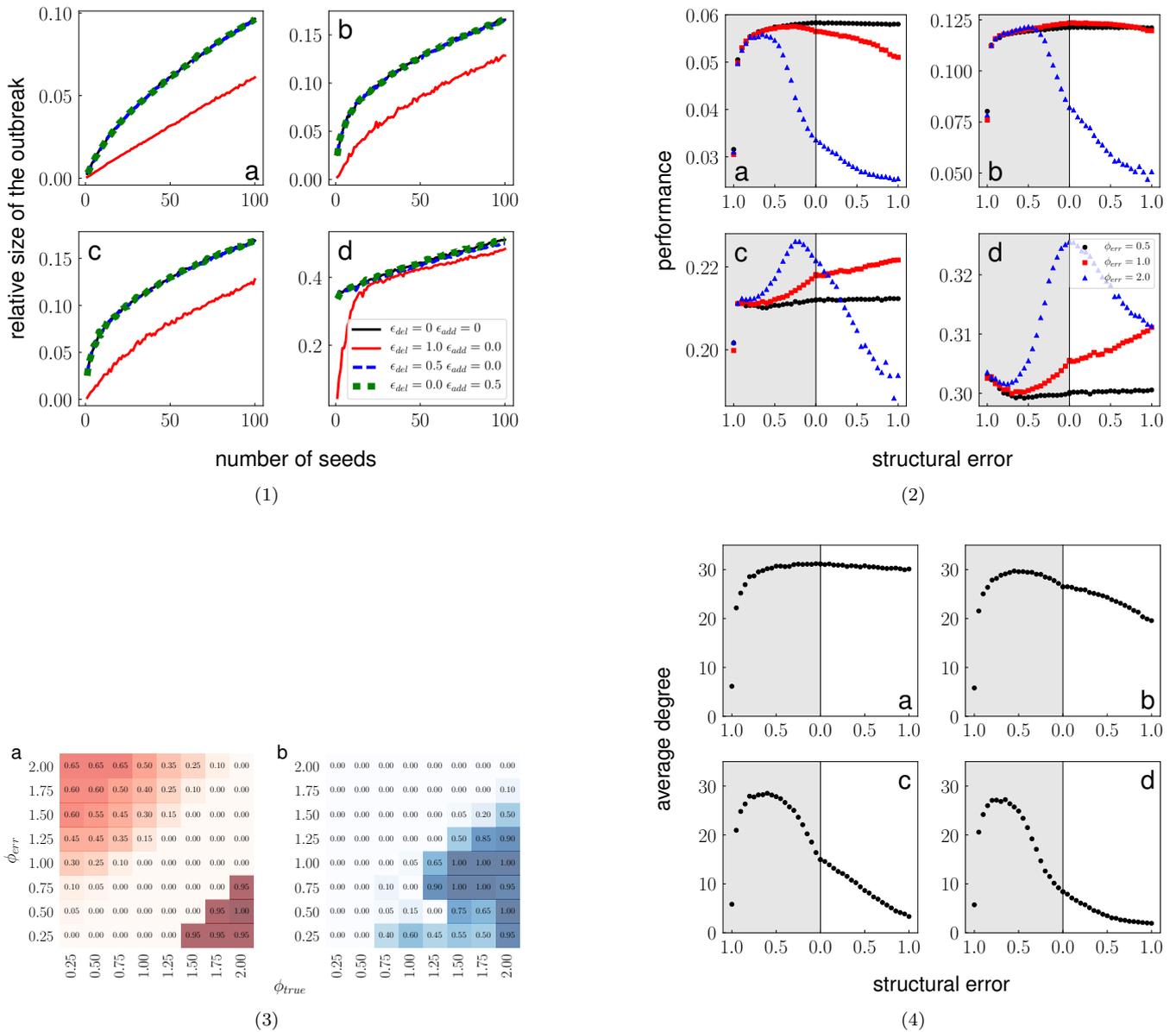

(1)

(2)

(3)

(4)

Figure S26: Yeast, protein. $|Q_{err}| = 100$

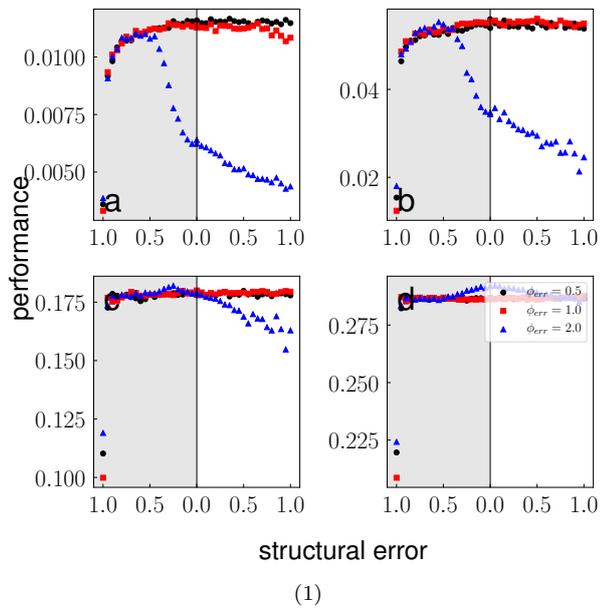

(1)

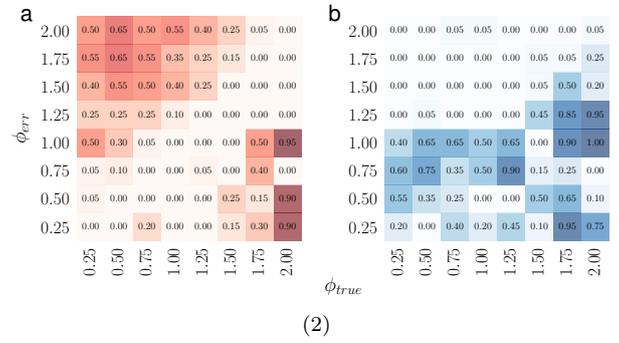

(2)

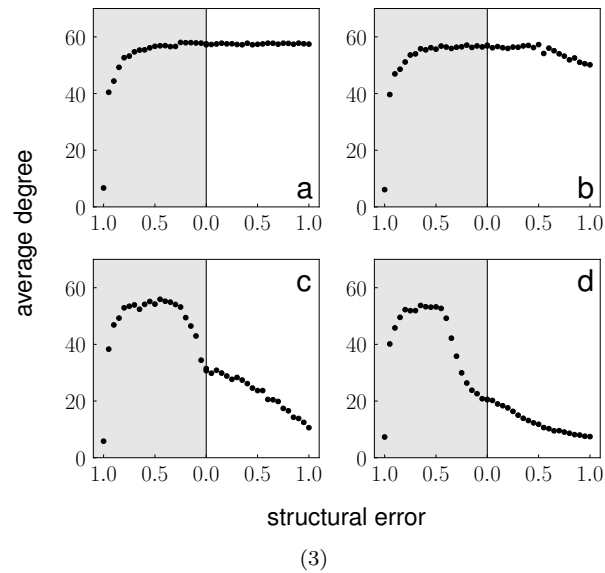

(3)

Figure S27: Yeast, protein. $|Q_{err}| = 10$



\end{document}